\documentclass[nhess, manuscript]{copernicus}

\nolinenumbers

\newcommand{\elev}{\text{elev}}

\newcommand{\bs}{\textbf{s}}

\newcommand{\bone}{\textbf{1}}

\begin{document}

\title{Model-based tools for assessing space and time change in daily maximum temperature: an application to the Ebro basin in Spain}

\Author[1]{Ana C.}{ Cebri\'an}

\Author[1]{Jes\'us}{ As\'in}

\Author[1]{Jorge}{Castillo-Mateo}

\Author[2]{Alan E.}{Gelfand}

\Author[1]{Jes\'us}{Abaurrea}

\affil[1]{Department of Statistical Methods, University of Zaragoza, Pedro Cerbuna 12, Zaragoza 50009, Spain}

\affil[2]{Department of Statistical Science, Duke University, Old Chemistry Building, Durham NC 27710, USA}

\correspondence{Ana C. Cebrián  (acebrian@unizar.es)}

\runningtitle{Model-based tools for assessing space and time change in daily temperature}

\runningauthor{TEXT}

\received{}
\pubdiscuss{} 
\revised{}
\accepted{}
\published{}

\firstpage{1}

\maketitle

\begin{abstract}
There is continuing interest in the investigation of change in temperature over space and time.  
We offer a set of tools to illuminate such change temporally, at desired temporal resolution, and spatially, according to region of interest, using   data  generated from  suitable space-time  models.  These tools include predictive spatial probability surfaces and spatial extents for an event.  Working with exceedance events around the center of the temperature distribution, the probability surfaces capture the spatial variation in the risk of an exceedance event, while the spatial extents capture the expected proportion of incidence of a given exceedance event for a region of interest. Importantly, the proposed tools can be used with the output from any suitable model fitted to any set of spatially referenced time series data.
 As an illustration, we employ a dataset from 1956 to 2015 collected at 18 stations over Arag\'{o}n in Spain, and a collection of daily maximum temperature series  obtained  from posterior predictive simulation of a Bayesian hierarchical daily temperature model.
The  results  for the summer period show that although there is an increasing risk in all the events used to quantify the effects of climate change, it is not spatially homogeneous, with   the largest increase arising  in   the center of Ebro valley and  Eastern Pyrenees area. The risk of an increase of the average temperature  between 1966-1975 and 2006-2015  higher than $1^\circ$C is  higher than 0.5 all over the region, and close to 1  in the previous areas. The extent  of daily temperature  higher than the reference  mean has  increased  3.5\% per decade. The mean of the extent indicates  that 95\% of the area under study has suffered a positive increment  of the average temperature, and almost 70\% higher than $1^{\circ}$C.
\end{abstract}

\introduction

Climate change is a global phenomenon, and the interest in assessing global warming in a spatio-temporal framework is  clear. However,  studies  to assess  and quantify the trends and  effects of climate change on temperature usually focus on the  study of  areally  aggregated signals 
or on the individual study of  local time series. 
Individual study  is limiting, because it does not allow us to assess the nature of changes  that may occur over a spatial region of interest. Further, studying spatially aggregated data sacrifices insight into local variation in behavior.
Concerning time scale, many spatial analysis   model annual or seasonal summaries of temperature, see \cite{masson2021climate} for a review.
However,   the use of a daily scale is important, since it  allows us to incorporate the inherent variability of  data while still enabling aggregation to a desired broader time scale. This scale is also essential to study persistence of temperatures. In addition, many  environmental applications  require  temperature data at this scale.  

Assessment of  space and time changes in daily temperature using empirical approaches has many limitations, particularly the inability to assess uncertainty. \cite{Dowlatabadi93} noted that uncertainty consideration should be an integral part of the integrated assessment of climate change. 
Also, \cite{Katz02} strongly advised  the use of full-fledged uncertainty analysis as part of climate assessment, recommending probabilistic modeling and, in particular, Bayesian hierarchical modeling and MCMC simulation techniques.
By now many  space-time environmental science models  have been proposed in a Bayesian framework \citep{Angulo98, Hartfield03,Craigmile11}. These Bayesian models are developed in a hierarchical form, at point level, introducing spatial random effects as a process model to capture the spatial correlation over the study region along with pure error terms to capture the uncertainty associated with the data relative to the process model \cite[see, e.g.,][]{Thorarinsdottir2017}. Other Bayesian models  pursue their analysis on a grid, such as \cite{Stroud01} and \cite{Wikle01}. \cite{Castillo2022} presented a point-referenced hierarchical model   for daily temperatures, which will be used as illustration in this work.

The contribution of this  paper is the proposal of tools to analyze in a space-time framework the evolution of daily  temperatures in a region  using   data  generated from   Bayesian space-time  models or suitable model-based stochastic weather generators (SWG).    The suggested tools cannot be applied to the output of an arbitrary SWG.  The SWG must enable simultaneous predictive generation of series for arbitrary unobserved sites \citep{Wilks99, Wilks09, Caraway14, Smith18}.   More precisely, the required predictive data is  a collection (replicates) of daily temperature series at a fine grid of geo-coded locations in the region under study. Some SWGs providing this type of data are based on Bayesian models, see for example  
\cite{Kleiber13} or \cite{verdin2019baygen}.
Here, we  will use data generated from the  model  by \cite{Castillo2022} as an illustration, but  the proposed approach can be used with predictive replications generated from any suitable model for any set of spatially referenced time series including, e.g., precipitation data.

We offer two main strategies for quantifying  the effect of climate change on different features of temperature, each with associated uncertainty. The first calculates probabilities that will be useful in climate risk assessment.  In fact, according  to \cite{Katz02}, the quantification of uncertainty in the form of probabilities is required as input to any decision or risk analysis. United Nations Framework Convention on Climate Change (UNFCCC) defines climate risk as the probability of exceeding one or more criteria of vulnerability. The approach suggested in this work aims to compute the occurrence probabilities of this type of event, defined in terms of climate signals exceeding a threshold. 
Further, this approach enables calculation of probability surface maps for the events of interest, capturing the spatial behavior of the probability of a given event.

The second strategy formalizes the concept of an extent   to investigate a useful objective in spatial analysis of climate, i.e., to characterize the extent of occurrence of a specific feature within a given area. More precisely, the extent associated with a given region reflects the proportion of the region in which the event is expected to occur. We can specify this at daily scale but further, we can average over days to attach this inference to coarser time scales.   Using this approach, we are able  not only to identify the areas where  a feature of interest occurs but also to quantify the  mean  and  uncertainty of the percentage  of area where that feature occurs. 
There are  previous studies analyzing  the idea of extent of extreme temperatures using observed data \citep{rebetez2009analysis,keellings2020spatiotemporal} or climate model output \citep{khan2019trends,lyon2019projected}. 
However, they  employed descriptive approaches precluding formal inference. Some formal concepts related to the notion of an extent have been introduced in the statistical literature.  \cite{Bolin15} and \cite{Sommerfeld18} consider excursion sets, which are sets of points in an area where a spatial function is above a given threshold.   \cite{haug2020spatial} identified  excursion sets  in Europe with significant trends in summer mean temperature. \cite{cebrian2021spatio} defined the notion of the extent  of  an extreme heat event as a stochastic object and  used it to calculate daily, seasonal and decadal averages.  
Excursion sets and level sets are examples (we consider others) of local events whose proportion of incidence, i.e., prevalence over a subregion of interest, enables greater insight into temperature behavior.

To show the applicability of our tools,   we consider events  having a  temperature  higher than the  corresponding local mean, or an increase in the mean temperature between two decades higher than a given value. In addition, any other event defined in terms of the available time series and a specified threshold can be considered.
Using the  proposed strategies,  we compute the probabilities of a positive increment of temperature  between two decades.
Further, we characterize, for a given day within a given year,  what proportion of the subregion was above a choice of a local reference temperature during one day or during a run of consecutive days, in order to study persistent temperatures.  Moreover, we study the behavior of these extents over time and also comparatively between subregions.  Since our generating  model is autoregressive,  correlation structure in the series is captured and we can use our tools to formally investigate persistence.  More precisely, we can study runs  of days with the same  climate event, which is a common approach to study this feature     \citep{pfleiderer2018quantification, tye2019climate}. These persistent  events are particular cases of the compound events defined by \cite{zscheischler2020typology}. The importance of the  study of the effects of climate change on temperature persistence is underscored by \cite{Li21}.

The proposed tools are employed to analyze  temperature evolution in an area around Aragón (Spain).  The tools are applied to a collection of posterior predictive gridded daily temperature series   obtained using output from the point-referenced hierarchical model by \cite{Castillo2022}. This model was fitted  using observed daily maximum temperatures at $n = 18$ sites, from 1956 to 2015. It is a rich autoregressive mean model which captures needed spatial dependence through four Gaussian processes (GPs) modeling  intercepts, slope/trend coefficients, variances, and autocorrelations, respectively. A brief summary of the model is given in Section 2.  While alternative models could be proposed, the model we employ  was  validated for this dataset in \cite{Castillo2022} to reproduce the statistical properties of the central part of the daily temperature distribution.

The paper is structured as follows. Section~\ref{S2} describes  the  observed temperature series  and the space-time  model  by \cite{Castillo2022} used to generate the grid of simulation replicates of temperature series. Section~\ref{S3} presents the proposed tools  for the space-time analysis  of  the replications of the temperature series. Section~\ref{S4}  summarizes the results of the analysis of two types of events, those  based on the comparison of temperature with a reference value, and those based on the  temperature increments between two decades. It also shows the comparison of the evolution of  the extent in two areas with different climates.
Finally, Section~\ref{S5} summarizes the  main conclusions and future work.

\section{The dataset and the model}
\label{S2}

Here, in Section 2.1 we present the dataset used and in Section 2.2 the model fitted to it. 

\subsection{The dataset and some exploratory analysis}

\begin{figure}
	\begin{center}
		\includegraphics[trim=2cm 2cm 2cm 1cm,width=0.45\linewidth]{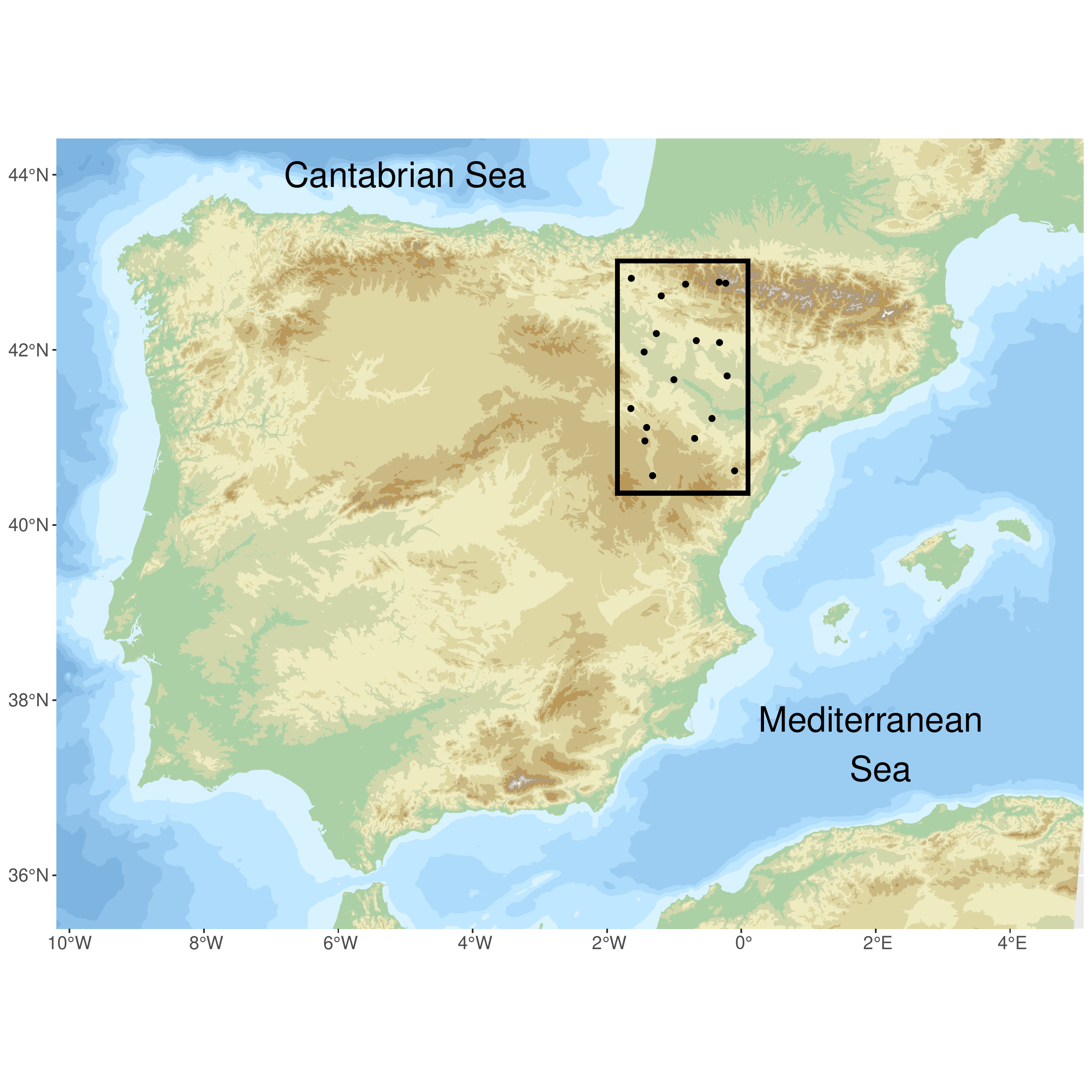}
		\includegraphics[trim=3cm 2cm 3cm 10cm, width=0.15\linewidth]{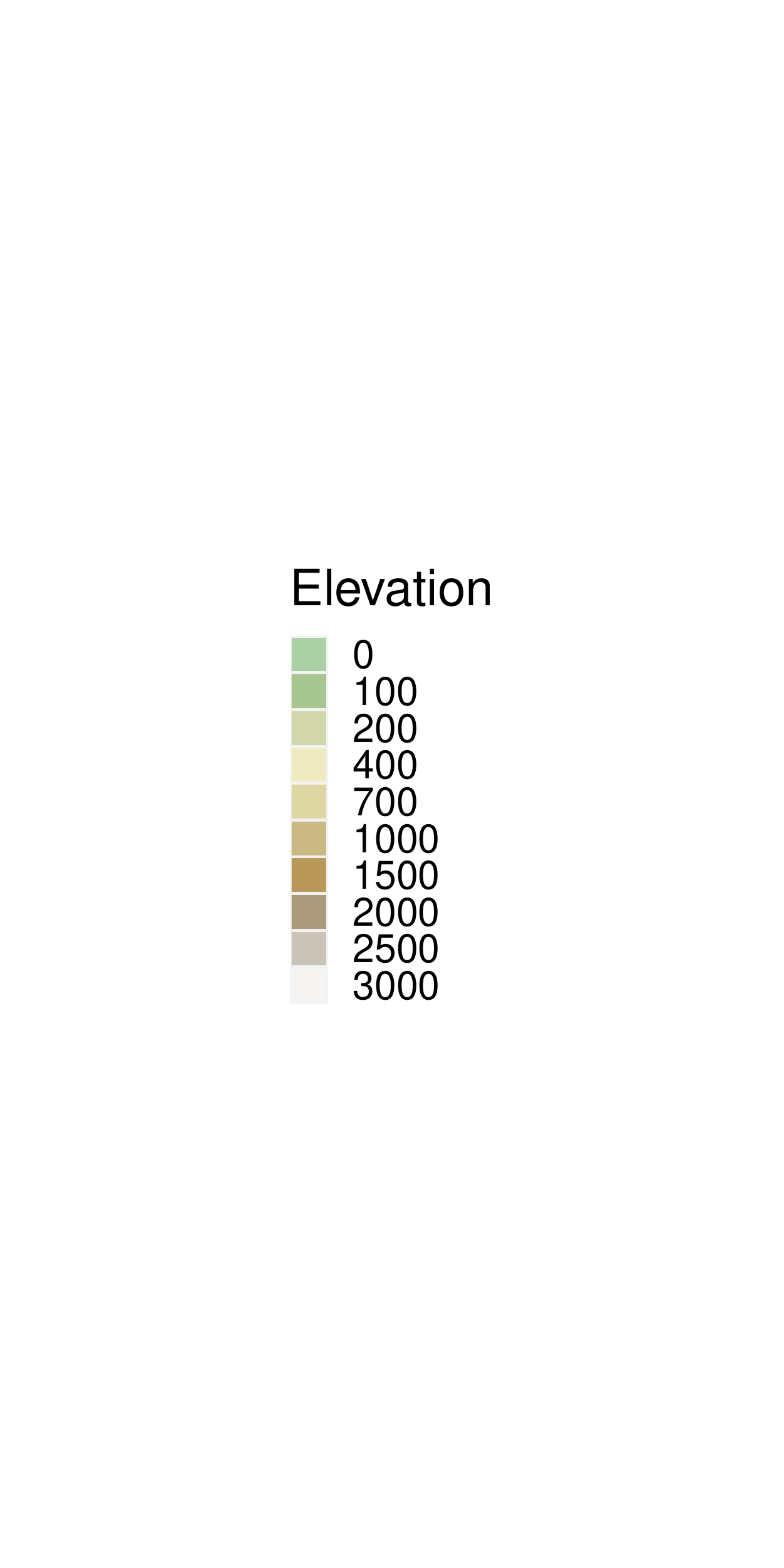}
		\includegraphics[trim=4cm 0cm 0cm 0cm, width=0.35\linewidth]{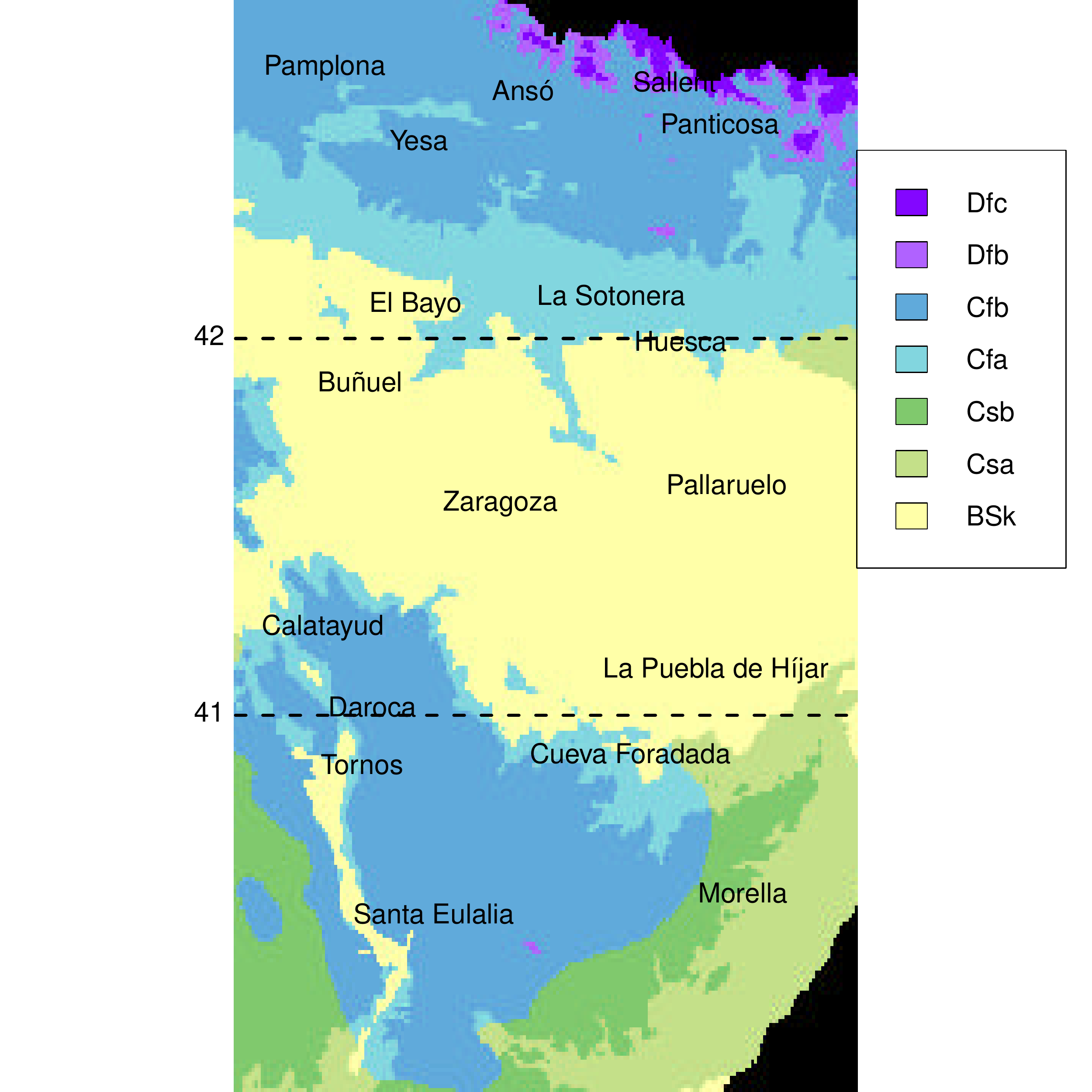}
		\caption{Right: Relief map  of the region under study and  location of observed temperature series. Left: Climate classification of the region.}
		\label{Fig_relief_climate}
	\end{center}
\end{figure}

The study area is located in the Ebro basin (85,362 km$^2$), in the northeast of Spain, see Fig.~\ref{Fig_relief_climate}. Different climate subareas can be distinguished, 
due to its location in the Iberian Peninsula and its heterogeneous orography that includes the Ebro valley (center) where elevations descend to 200 m, and mountains:  Pyrenees (north), Cantabrian Range (northwest) and Iberian System (southwest). The mountains reach 3,000 m in the Pyrenees and 2,000 m in the Iberian System. Mediterranean-continental dry climate with irregular rainfall and a large temperature range is the prevailing climate, but also mountain climates are present in the region. This variety of climate conditions is one reason for interest in the area.

Figure \ref{Fig_relief_climate} shows the location of the $18$ sites where daily maximum temperature observed series,  from 1956 to 2015, are available. They have been provided by the Spanish Meteorological Office (AEMET). 
Temperature in this region shows seasonal behavior, with large  differences between winter and summer months; e.g.,  in Zaragoza (the main city in the region)  this  difference  is around 22$^{\circ}\text{C}$.  This seasonal  pattern  is quite spatially homogeneous in the area.

\cite{Castillo2022} analyzed  the warm  period between May 1st and September 30th using these temperature series, and fitted the model described in Section \ref{Smodel} to them. A thorough exploratory analysis of these series  can be found in that work, and a summary in  the Supplement, Section S1. 
According to this exploratory analysis, spatial variability in the mean temperatures is linked to elevation, where Panticosa is the highest and coldest location and La Puebla de Híjar in the  valley is the hottest.  However, elevation is not sufficient to explain the mean temperature variability, since there are areas  at the south and north of the Ebro river, with similar elevation, around 1,000 m, and different mean temperatures. The  standard deviations of the series show the maximum  variability is in the northwest, 5.6$^{\circ}\text{C}$ in  Pamplona, and the minimum is in the southwest, 4.1$^{\circ}\text{C}$ in Cueva Foradada. The serial correlation is over 0.90 for all series, reflecting temperature inertia in the short term.  It is a key distributional feature to be considered in presenting the statistical inference. 
To explore the observed change over time, linear trends are estimated  in each series using the observation in the period JJA 1956-2015. Spatially heterogeneous behavior is found in this feature, with the smallest changes in the western observatories and the largest  in the valley.

With regard to the spatial dependence  between the daily temperature series,  a strong correlation between them is observed, and therefore, should be incorporated into the analysis. The pairwise Pearson coefficients  are calculated separately  for each month to avoid  the correlation caused by the common seasonal pattern. It is found that the 25th percentile of those coefficients is 0.82 in June, 0.74 in July and 0.73 in August.

\begin{figure}[t]
	\begin{center}
		\includegraphics[width=0.4\textwidth]{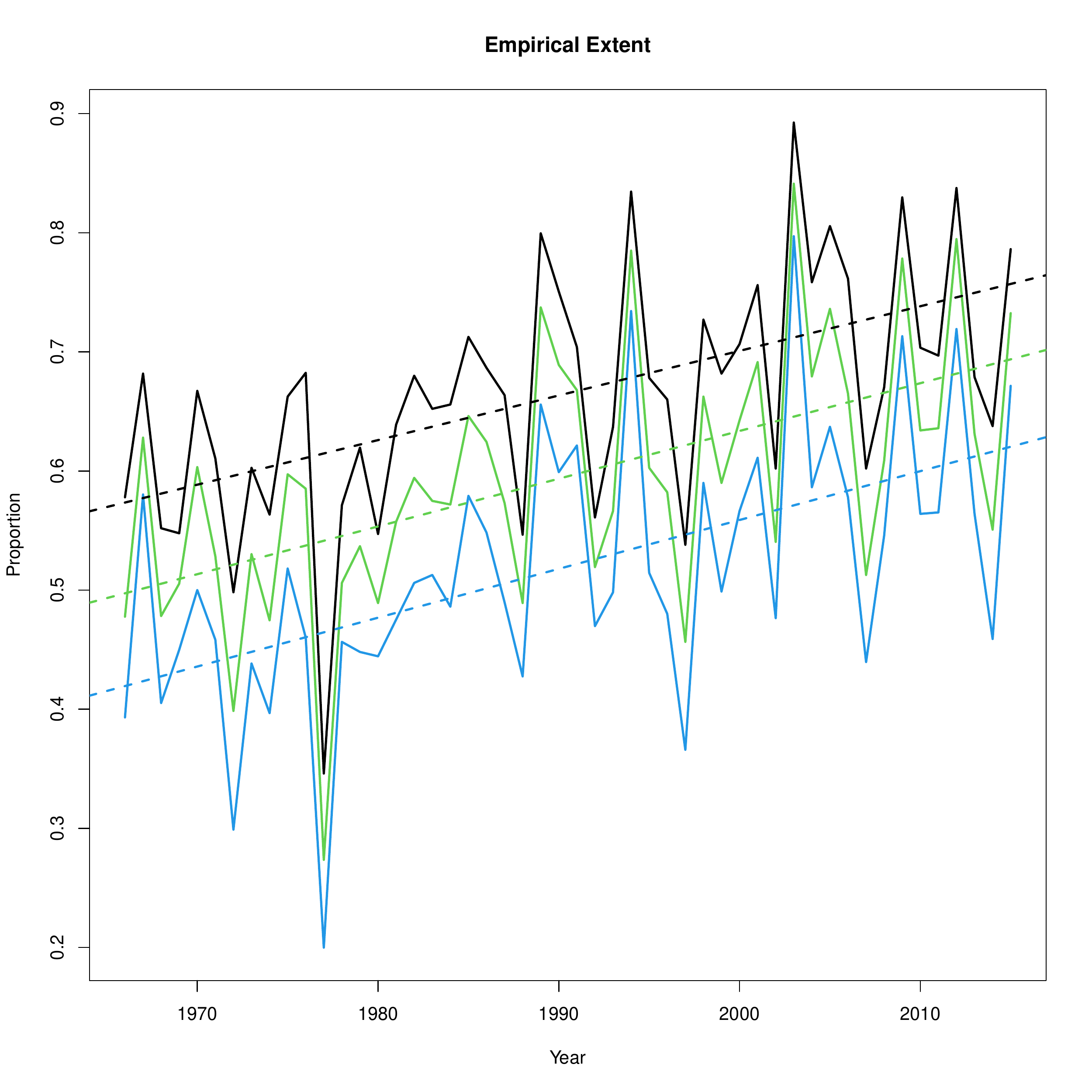}	
	\end{center}
	\caption{Yearly averages in JJA  of  the empirical extent  for increments over the reference mean $\tilde \mu(\bs)$  higher than  $0$ (black), $1$ (green) and  $2^\circ$C (blue) and linear trends fitted over time.}	 \label{FigExplor_Extents} 
\end{figure}

Finally, to explore  changes over time  and space, we consider  the ``empirical extent" for the event defined  as the  increment of daily temperature above a reference mean $\tilde \mu (\bs)$ higher than a value $c$.  The empirical extent is computed as  the observed proportion of  the 18 available stations  where the event occurs.  Figure \ref{FigExplor_Extents} summarizes the  average of the empirical extent  over days in JJA of each year during the period 1966-2016  for  events based on  increments over the reference mean  $\tilde \mu(\bs)$ higher than $c= 0, 1$ and $2^\circ$C. $\tilde \mu(\bs)$ is a reference mean that is site-specific but constant over time;  details of its definition  can be found  in Section \ref{S4}. The fitted linear trend  shows an increase of the empirical extent of  $0.037$ per decade for increases over $\tilde \mu(\bs)$   higher than $0$ and  $0.041$ for  increases higher than $2^\circ$C.   
An evident limitation of this empirical extent  is that it is  based on only 18 stations. 

\subsection{A space-time model for daily  temperatures in the warm period}

\label{Smodel}

The following spatio-temporal model, fitted and validated in \cite{Castillo2022}, has been used as illustration to generate the posterior predictive realizations of the time series that are employed with the tools presented in Section 3.  As noted above, alternative models  able to generate adequate replicates of time series at a fine grid of geo-coded locations could equally well be used to generate the required data to use the proposed tools; e.g. the  Bayesian space-time model by \cite{schliep2021long} or the SWG by   \cite{verdin2019baygen}

An exploratory analysis of the temperature dataset shows that an adequate statistical model for daily temperatures must include terms that capture the seasonal behavior, the spatially heterogeneous standard deviation and trend, as well as the temporal and spatial dependence observed in the region. As a result, each observation will be indexed by a location in Aragón, a year, and a day during the warm period.  To obtain the desired behavior, we introduce both fixed effects terms and random effects terms. The fixed effects can capture elevation as well as trend and seasonal patterns.  The random effects can capture spatial and temporal dependence structure as well as providing surrogates for unobserved or unmeasured spatial or temporal regressors.

The Bayesian hierarchical model by \cite{Castillo2022} incorporates all these requirements. The daily maximum temperature for day within year $\ell$ (for warm period from May to September), year $t$, and location $\bs$, $Y_{t,\ell}(\bs)$, is modeled as
\begin{eqnarray}   \label{eq:model}
	Y_{t,\ell}(\bs) & =& m_{t,\ell}(\bs) + \rho(\bs) \left(Y_{t,\ell-1}(\bs) - m_{t,\ell-1}(\bs)\right) +  \epsilon_{t,\ell}(\bs), \\	 \notag
	m_{t,\ell}(\bs)& = & \mu_{t,\ell}(\bs) + \gamma_t(\bs), \\ \notag
	\mu_{t,\ell}(\bs)& = &\beta_0 + \alpha t + \beta_1 \sin(2 \pi \ell / 365) + 
	\beta_2 \cos(2 \pi \ell / 365) + \beta_3 \elev(\bs), \\ \notag
	\gamma_t(\bs)& = &\beta_0(\bs) + \alpha(\bs) t + \psi_t + \eta_t(\bs).
\end{eqnarray}

The model reveals the hierarchical structure, modeling the data given a mean and then modeling the mean.  It introduces temporal dependence  using a first-order autoregressive  structure on the temperature anomalies, as suggested 
in the Fifth IPCC Report   \citep{hartmann2013climate}. Then, $\rho(\bs)$ is a spatially varying autoregression coefficient that captures the serial correlation for consecutive days at location $\bs$. The conditional mean of $Y_{t,\ell}(\bs)$ given yesterday's temperature $Y_{t,\ell-1}(\bs)$ is expressed by $m_{t,\ell}(\bs) + \rho(\bs) \left(Y_{t,\ell-1}(\bs) - m_{t,\ell-1}(\bs)\right)$. Modeling of the serial correlation in the data is important since its omission  may lead to an inappropriate  statistical assessment  of the trend \citep{zwiers1995taking, scott2011statistical}.
The model assumes that spatial and temporal dependence is captured by the conditional mean, so that  $\epsilon_{t,\ell}(\bs)$ are pure error terms with  independent $N(0,\sigma^2(\bs))$ distribution where $\sigma^2(\bs)$ is a spatially varying  variance.

Here, $m_{t,\ell}(\bs)$ contains  fixed and random effects, $\mu_{t,\ell}(\bs)$ and $\gamma_t(\bs)$, respectively. The daily fixed effects are captured by $\beta_0$, a global intercept, $\alpha t$, a baseline long-term linear trend, $\beta_1$ and $\beta_2$, the coefficients of a harmonic that captures the seasonal component within the $153$ day warm period, and $\beta_3$, the coefficient for  the elevation at $\bs$, $\elev(\bs)$.

The annual random effects given in $\gamma_t(\bs)$ capture space-time dependence through GPs \citep{BCG}.   A local spatial adjustment to the intercept, $\beta_0(\bs)$, and a local slope adjustment, $\alpha(\bs)$, enable a flexible, spatially varying, local linear trend.  This \emph{locally} linear trend substantially extends the usual linear trend specification adopted in climate analysis \citep{masson2021climate}. The terms  $\psi_t \sim \text{IID } N(0, \sigma_{\psi}^{2})$ provide annual intercepts to allow for yearly shifts (associated, e.g., with the ENSO), and $\eta_t(\bs) \sim \text{IID } N(0, \sigma_{\eta}^{2})$ provides local annual intercepts to allow for local yearly shifts.

Thus, four GPs are introduced. First, $\beta_0(\bs)$ and $\alpha(\bs)$ are GPs with zero mean and exponential covariance function.   We specify $\rho(\bs)$  using the customary transformation to the range of correlation, through the GP $Z_{\rho}(\bs) = \log\{(1 + \rho(\bs)) / (1 - \rho(\bs))\}$ with mean $Z_{\rho}$ and exponential covariance function. Similarly, we specify the positive $\sigma^2(\bs)$   through the GP $Z_{\sigma}(\bs) = \log\{\sigma^2(\bs)\}$ with mean $Z_{\sigma}$ and exponential covariance function. For more detail on GPs see, e.g., Chapter~3 in \cite{BCG}.

The model is fitted in a Bayesian framework using MCMC \citep{BCG}.  Additional information regarding prior specification is available in the Supplement, Section S2.
\cite{Castillo2022} offer computational details of a Gibbs sampler algorithm  \citep{Gelfand90} for model fitting. 

Posterior samples of model parameters are used to obtain posterior predictive replicates of temperature series over a regular grid of geo-coded locations in the region, using the posterior predictive distribution. Inference for the parameters in model \eqref{eq:model}  can be implemented using samples  obtained from the resulting  joint posterior probability distribution.  As a last comment, the result of the Bayesian model fitting is to produce the posterior distribution of any unknown in the model, i.e., the conditional distribution of the unknown given the data.  Using MCMC to fit the model enables as many samples as we wish from this posterior distribution.  From these samples, we can learn arbitrarily well about any features of the distribution of the unknown including say, the mean and variance, as well as interval estimates. 

\subsubsection{Dataset generated from the model}

As noted above, the model in \eqref{eq:model} enables kriging to unobserved locations for a given year $t$ and day within year $\ell$.  At a new subset of sites we employ composition sampling \cite[Ch.~6]{BCG}  to obtain a sample from the joint posterior predictive distribution of daily maximum temperatures for  any time $(t,\ell)$. Briefly, the idea of composition sampling is to obtain posterior predictive replications from  model \eqref{eq:model}, using a sample of the parameters, GP replications, and errors. Posterior samples for the parameters are available from the model fitting.  Joint posterior samples for the GPs are obtained using posterior samples of the parameters, through usual Bayesian kriging, and  for the errors  using posterior samples of the spatially varying variances, by simulating normal random variables. 

Altogether, a collection of independent replicates, of daily temperature within a year on a spatial grid $\mathcal{D}$ for the  period of interest, $\{Y^{(b)}_{t,\ell}(\bs);\, b=1,\ldots, B\}$, can be generated.  Note that $B$ can be as large as we wish; it has no connection to the size of the dataset used to fit the model.  As a sample from a predictive distribution, 
these  replicates provide the fundamental material for all of the inference using the tools in the sequel. They will not only allow us to learn about the distribution of temperature at any location on any day but also about the distribution of any other measures of interest that can be computed as functions of  temperatures $Y_{t,\ell}(\bs)$,  as we describe  in the following sections.  

We emphasize again that this posterior predictive approach can be implemented  using  datasets from other generating models.  That is, for other datasets, over different regions and appropriate time scales, fitted with different appropriate models, we can follow the same path for enhanced learning about temperature behavior over space and time.

\section{Novel tools  for enriching space-time analysis}
\label{S3}

We present tools to illuminate the spatial and temporal behavior of daily maximum temperature. Again, the set $\{Y^{(b)}_{t,\ell}(\bs);\, b=1,\ldots, B\}$ provides samples of any function of daily temperature over days, years or locations.  Hence, we can ``see'' the distribution of this function and any features of this distribution that are of interest such as its center and variability.

A primary intent is to study changes in temperature over time, to quantify their  magnitude,  and to identify  areas with different evolution.   To obtain conclusions over space, we use probability (risk) surfaces and the concept of extent (proportion of area)   linked to ``events'' that allow the quantification of the increase in temperature. The underlying idea is to define events of interest $A_{t,\ell}(\bs)$ in terms of the daily temperatures $Y_{t,\ell}(\bs)$, e.g.,  the event of  temperature at day $(t,\ell)$  at location $\bs$ being higher than a site-specific reference value $r(\bs)$, $A_{t,\ell}(\bs)=\{Y_{t,\ell}(\bs)>r(\bs)\}$. From a model-generated replicate $Y^{(b)}_{t,\ell}(\bs)$,   we can obtain a realization of the binary/indicator variable,   $\bone(Y_{t,\ell}(\bs)>r(\bs))$, a variable that  is equal to 1  if $Y_{t,\ell}(\bs)>r(\bs)$, and 0 otherwise.  For illustration, all the measures and tools in this section are defined for  the simple event $\{Y_{t,\ell}(\bs)>r(\bs)\}$.  However, they can be applied to any other event defined in terms of $Y_{t,\ell}(\bs)$, such as those  introduced in Section \ref{SDefEvents}.  All of the ensuing inference is posterior, i.e., conditional given the data. To simplify notation, we suppress the conditioning below, then $P(A_{t,\ell}(\bs) | data)$ will be denoted $P(A_{t,\ell}(\bs))$.

\subsection{Posterior probability surfaces}

The posterior probability associated with an event  at location $\bs$, $A_{t,\ell}(\bs)$, is obtained by calculating  the proportion of events in the  collection of realizations $Y^{(b)}_{t,\ell}(\bs)$, $b=1,\ldots,B$, i.e., the mean of the binary variables indicating the occurrence of the event,     

\begin{eqnarray}
	\hat P\left(A_{t,\ell}(\bs)\right) = \frac{1}{B}
	\sum_{b=1}^B \bone \left(A^{(b)}_{t,\ell}(\bs)\right)
	\label{EqP}
\end{eqnarray}
where $A^{(b)}_{t,\ell}(\bs)$ is the event defined in terms of  the $b$ realization $Y^{(b)}_{t,\ell}(\bs)$; e.g.,  $\{Y^{(b)}_{t,\ell}(\bs)>r(\bs)\}$. 

Events based on daily temperature, such as $\{Y_{t,\ell}(\bs)>r(\bs)\}$,  are defined for each day $(t,\ell)$,  so that it is straightforward to summarize them over a period of time.
The previous daily probabilities can be summarized by averaging them in a given period, e.g.,  days in JJA in a decade $D$, denoted  D-JJA, 

\begin{equation}
	\bar{{P}}\left(A(\bs)\right)= \frac{1}{920} 	\sum_{t \in D, \ell \in JJA} \hat P\left(A_{t,\ell}(\bs)\right).
	\label{Eq0b}
\end{equation}

These daily or average probabilities over the grid of points $\bs$ can be  plotted and smoothed in a map, to reveal a probability surface, or averaged over a region.

\subsection{Extent for an event}

The extent for an event  in a region  $\mathcal{B} \subseteq \mathcal{D}$ is defined as the  proportion/fraction of incidence of that event in the region \citep{cebrian2021spatio}. Formally, the extent in $\mathcal{B}$ for an event $A_{t,\ell}(\bs)$ is the  integral, 
\begin{equation*}
	Ext\left(A_{t,\ell}(\mathcal{B}) \right) = \frac{1}{\|\mathcal{B}\|}\int_{\mathcal{B}} \bone \left(A_{t,\ell}(\bs)\right) d\bs
\end{equation*}
where $\|\mathcal{B}\|$ denotes the area of $\mathcal{B}$. Although this  integral cannot be calculated explicitly, it can be approximated arbitrarily well by Monte Carlo integration as 
\begin{equation}
	\widetilde{Ext}\left(A_{t,\ell}(\mathcal{B} )\right) = \sum_{\bs \in \mathcal{B}} w_{\bs} \bone\left(A_{t,\ell}(\bs)\right)
	\label{Eq2}
\end{equation}
where $w_{\bs}$  weights the  size of the grid cell  linked to $\bs$,  which cover region $\mathcal{B}$: $w_{\bs}=w_{\bs}^*/\sum_{\bs \in \mathcal{B}} w_{\bs}^*$ for given  size grid cell $w_{\bs}^*$.  In other words, it is the weighted average over the region  of the  binary variables  for events $A_{t,\ell}(\bs)$.  

We can obtain a realization of an extent from each set of realizations $Y^{(b)}_{t,\ell}(\bs)$ for $\bs \in \mathcal{B}$, and with $B$ observations of the extent, we obtain its posterior predictive distribution, which is employed for inference. 
To keep the notation simple, if the considered region is the entire region, $\mathcal{B}=\mathcal{D}$, the argument $\mathcal{B}$ is omitted.

When we compute daily extents, again, it may be of interest to summarize them   by averaging them over a period of time, e.g.,  D-JJA  with 920 ($10\times 92$) days,
\begin{equation} \label{Eq3}
    \overline{Ext}\left( A(\mathcal{B}) \right)=\frac{1}{920}\sum_{t\in D, \ell\in JJA}  \widetilde{Ext}\left( A_{t,\ell}(\mathcal{B}) \right).
\end{equation}

Note that the $B$  realizations  available of this average extent will characterize the distribution of the average, not  a daily extent. This means that the variance  will be much smaller than in the previous example since it is averaged over a large number of terms. 

\subsection{Defining events to quantify the increase in temperature}
\label{SDefEvents}

There are many ways to define events that allow us to quantify an increase in temperature. Here, we  propose several ways to define  those events, but any other option  that can be evaluated from the daily temperature observations $Y_{t,\ell}(\bs)$ can be  studied  by applying the tools described in the previous section.  We consider two  general choices of events, one based on increments over a reference value and the other  on increments between two periods of time.

First, we consider events defined in terms of the  increment in temperature with respect to a  reference value $r(\bs)$, which is site-specific but constant across time.  
The simplest  events,  $\{Y_{t,\ell}(\bs)-r(\bs)>c\}$, are based on daily temperature; note that these events correspond to events  defined as daily temperature higher than a value $r(\bs)+c$. An important feature of temperature is its persistence across days, so that we define events  based  on the daily temperature for $k=2$ or 3  consecutive days $\{Y_{t,\ell}(\bs) -r(\bs)>c;2\} \equiv \{Y_{t,\ell}(\bs)-r(\bs), Y_{t,\ell+1}(\bs)-r(\bs) > c\}$ or   $\{Y_{t,\ell}(\bs)-r(\bs) >c;3\} \equiv \{ Y_{t,\ell-1}(\bs)-r(\bs), Y_{t,\ell}(\bs)-r(\bs), Y_{t,\ell+1}(\bs)-r(\bs) >c\}$.

An extension  is to define events based on  an average temperature in a period of time, e.g.,  the average  in D-JJA,
$$\bar Y_{D}(\bs)=\frac{1}{920}\sum_ {t \in D, \ell \in JJA} Y_{t,\ell}(\bs).$$
Then, we define events based on the increment of the average temperature over the reference value, $\{\bar Y_{D}(\bs) -r(\bs)>c\}$.

Another important feature  to quantify global warming is the increment of temperature between two periods of time; here, we will consider the  increment between two decades 1966-1975 ($D1$) and 2006-2015 ($D5$). As above, the increments can be defined using daily temperatures or average temperatures.  Here we show the analysis of the increment of  average temperatures,  that is  the events $\{\bar Y_{D5}(\bs)-\bar Y_{D1}(\bs) >c\}$. The analysis of the increments between two decades  at a daily scale is presented in  the Supplement,  Section S4.1.  
 For clarity, Table \ref{TDef} summarizes the type and notation of all the events analyzed  in the following section.

\begin{table}[t]
	\begin{center}
		\caption{Events defined to quantify the effects of climate change.  In this work, events are defined for three values $c=0,1$ and $2^\circ$C and two persistence periods $k=2$ and 3 days, and the reference value $r(\bs)$ is  a local mean.}
		\label{TDef}
		\begin{tabular}{l|l}
			Event & Definition \\ \hline	
			$\{Y_{t,\ell}(\bs) -r(\bs)>c\}$ & Increment of daily temperature over a reference value $r(\bs)$, higher than $c$\\
			$\{Y_{t,\ell}(\bs) -r(\bs)>c;k\}$ &  Increment of daily temperature over a reference value $r(\bs)$,  higher than $c$ \\
			& in $k$ consecutive days  \\
			$\{\bar Y_{D}(\bs) -r(\bs)>c\}$ & Increment of average temperature in decade $D$ \\
			& over a reference  value $r(\bs)$ higher than $c$\\
			$\{\bar Y_{D5}(\bs)-\bar Y_{D1}(\bs) >c\}$ &  Increment of average temperatures between two decades \\
			& higher than $c$\\ 
		\end{tabular}
	\end{center}
\end{table}

\section{Results for the space-time analysis}
\label{S4}

We apply the methodology described in Section~\ref{S3} to study   the effect of climate  change  on different features related to daily temperature in the Ebro basin.  
Section \ref{S42} shows the results over the entire  region while a comparison of the extent for different increments of temperatures in  two areas  with different climates regimes   is  carried out in Section~\ref{S44}.

The tools are applied to  a set of $B=500$ replicates of daily temperature  $\{Y^{(b)}_{t,\ell}(\bs);\, b=1,\ldots,500\}$  generated from the model in Section~\ref{Smodel}, on a spatial grid covering the area $\mathcal{D}$  drawn  in Fig. \ref{Fig_relief_climate}, for the 92 days in JJA in the period 1956-2015. Given the  different orography in the study region, a grid with $4401$ points $\bs$ with a locally adapted spatial  resolution is adopted.  The  spatial changes in temperature  in flat areas are  slow so a $4\times 4$ km$^2$ grid is used, while in an area in the Pyrenees with a  steep relief, the scale of the grid is resolved to $1\times 1$ km$^2$.   The first decade of the generated realizations,  1956-1965, is reserved to obtain reference values and the analysis over time is done over the period 1966-2015.

\begin{figure}[t]
	\begin{center}
		\includegraphics[width=0.33\textwidth]{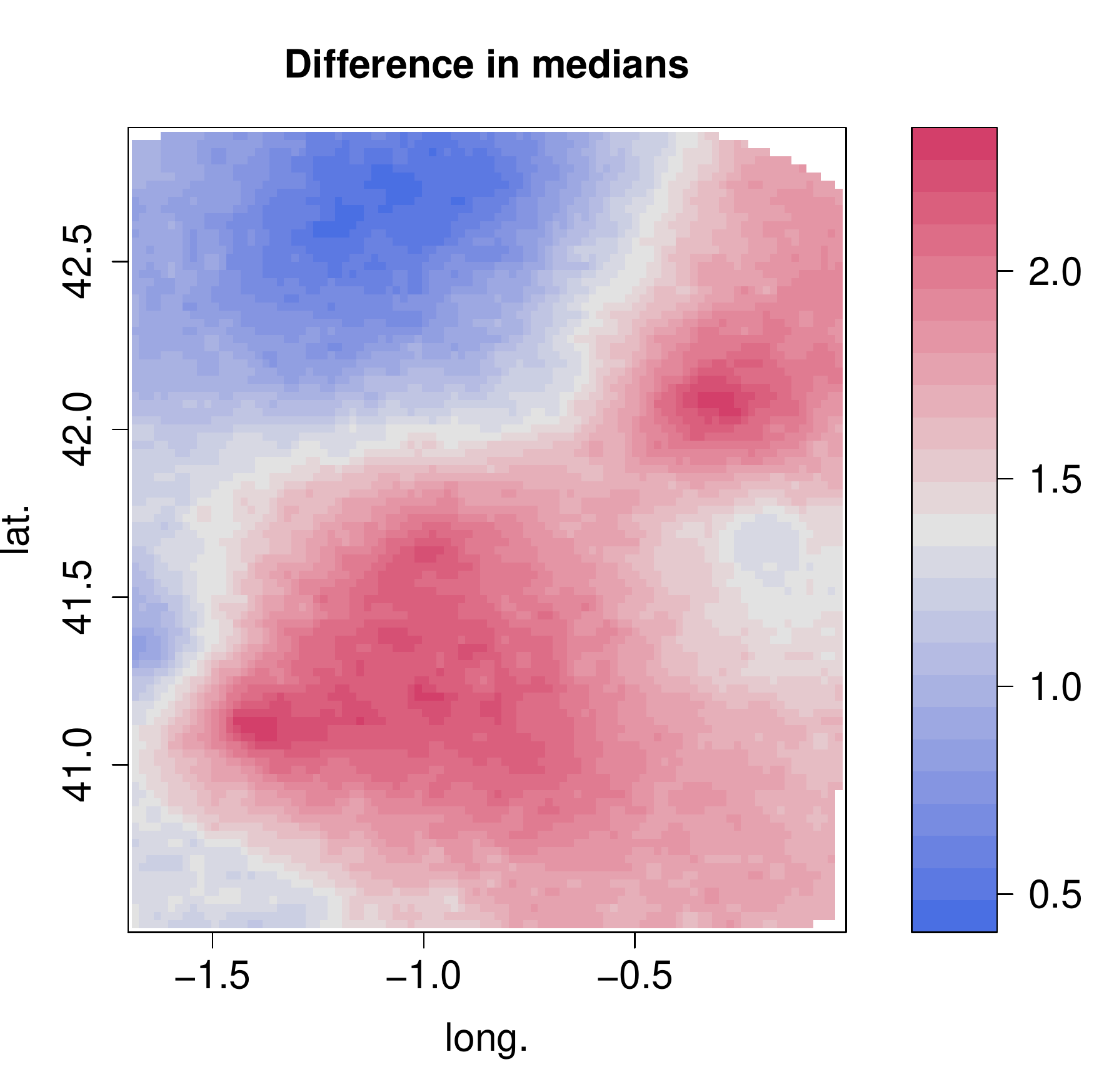}
		\includegraphics[width=0.33\textwidth]{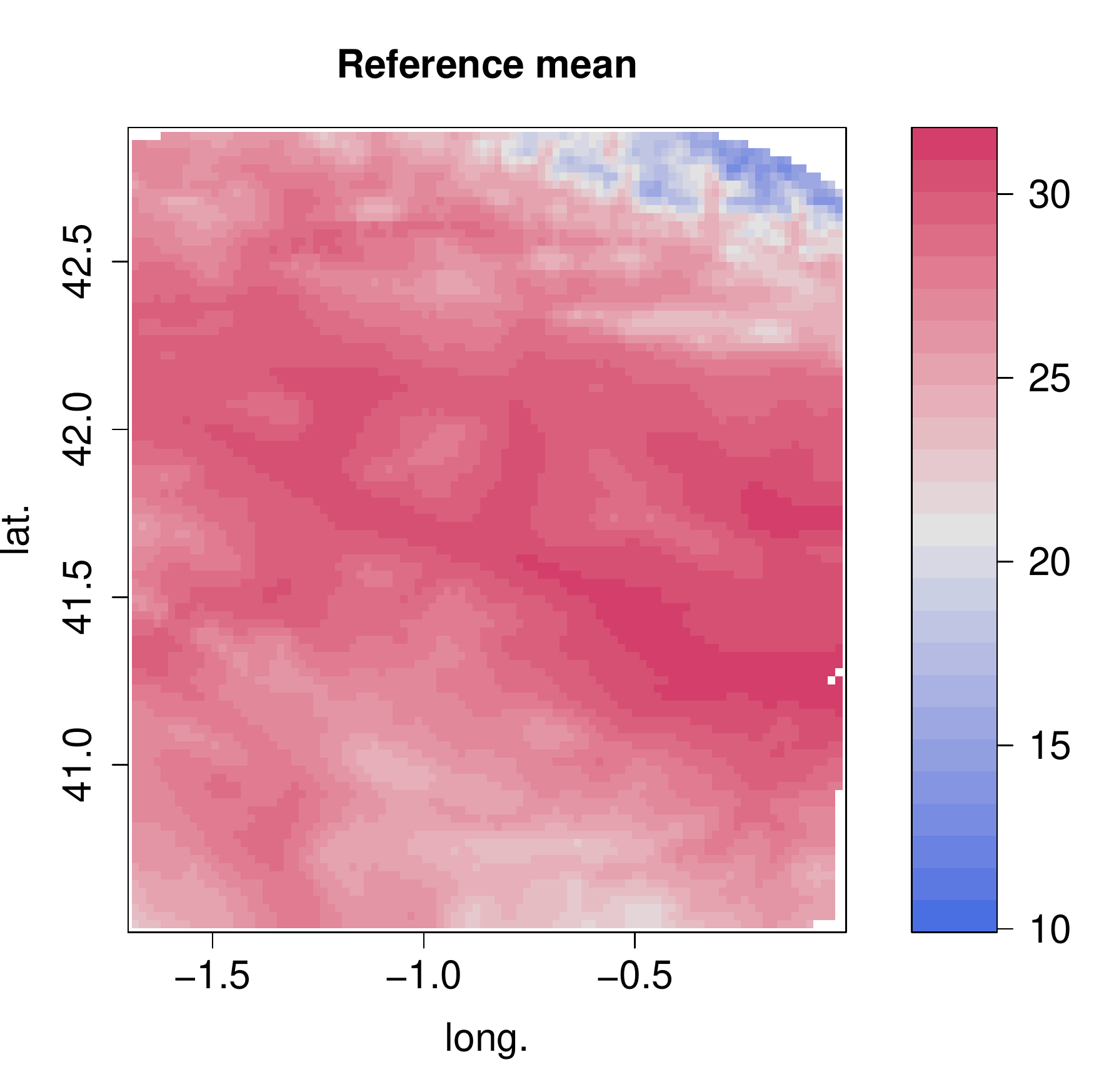}
	\end{center}
	\caption{Left: Difference in the medians of the daily temperatures ($^\circ$C) in JJA in decades $D5$ and $D1$. Right: Map of the reference mean $\tilde \mu(\bs)$, i.e. mean  daily temperature  ($^\circ$C) in JJA in the  decade 1956-1965.} \label{Fm}
\end{figure}

As a simple example  of the information provided by the output series, Fig. \ref{Fm} shows, spatially, the  difference between the medians  in decades  $D1$  and $D5$ (the medians in each decade are shown in  Fig. S2 in the  Supplement).  Although the increase is higher than 0.5$^\circ$C in all of the region, the map reflects the spatial variability of the area: the highest increases, greater than 2$^\circ$C occur in the center of the valley and  the east area of the  Pyrenees, while the lowest occur in the NW.

To define the first choice of events, we need a
reference value  $r(\bs)$. Here,   we consider a mean value,  but  other options, for example a high percentile,  could be used  to study the evolution of extreme events, provided that the considered data generator from the associated model is able to reproduce adequately the tails of data distribution.  Our site-specific  reference value $r(\bs)$ is the mean temperature in JJA during  the reference decade  1956-1965, denoted  as  the reference mean $\tilde \mu(\bs)$. 
The  resulting mean surface is shown in Fig. \ref{Fm} (right); the image is built  using the function \textit{pimage} from the library \textit{autoimage} \citep{French17}, that interpolates the previous points  on a regular grid using multilevel B-splines. 
The warmest area, with mean temperature higher than $30^\circ$C is the Ebro river valley, especially the areas closest to the river and the eastern  part of the valley, while the coolest areas  correspond to the Pyrenees, with mean temperatures lower than $20^\circ$C. We will analyze events for three different increments $c=0, 1$ and $2^\circ$C. The values  1 and 2 are approximately $1/4$ and $1/2$ of the standard deviation of  daily  temperature, and values in this range are commonly used to evaluate effects of climate warming  \citep{IPCC2018}.

\subsection{Analysis of the entire region}

\label{S42}

\subsubsection{Analysis of  increments of daily temperature over $\tilde \mu(\bs)$ }

This section summarizes  the results of the analysis of  events  based on increments of daily temperature over  the  reference mean, $\tilde \mu(\bs)$, for one day and persistent events for $k=2$ and 3 days.

\begin{figure}[tb]
	\begin{center}
		\includegraphics[width=0.31\textwidth]{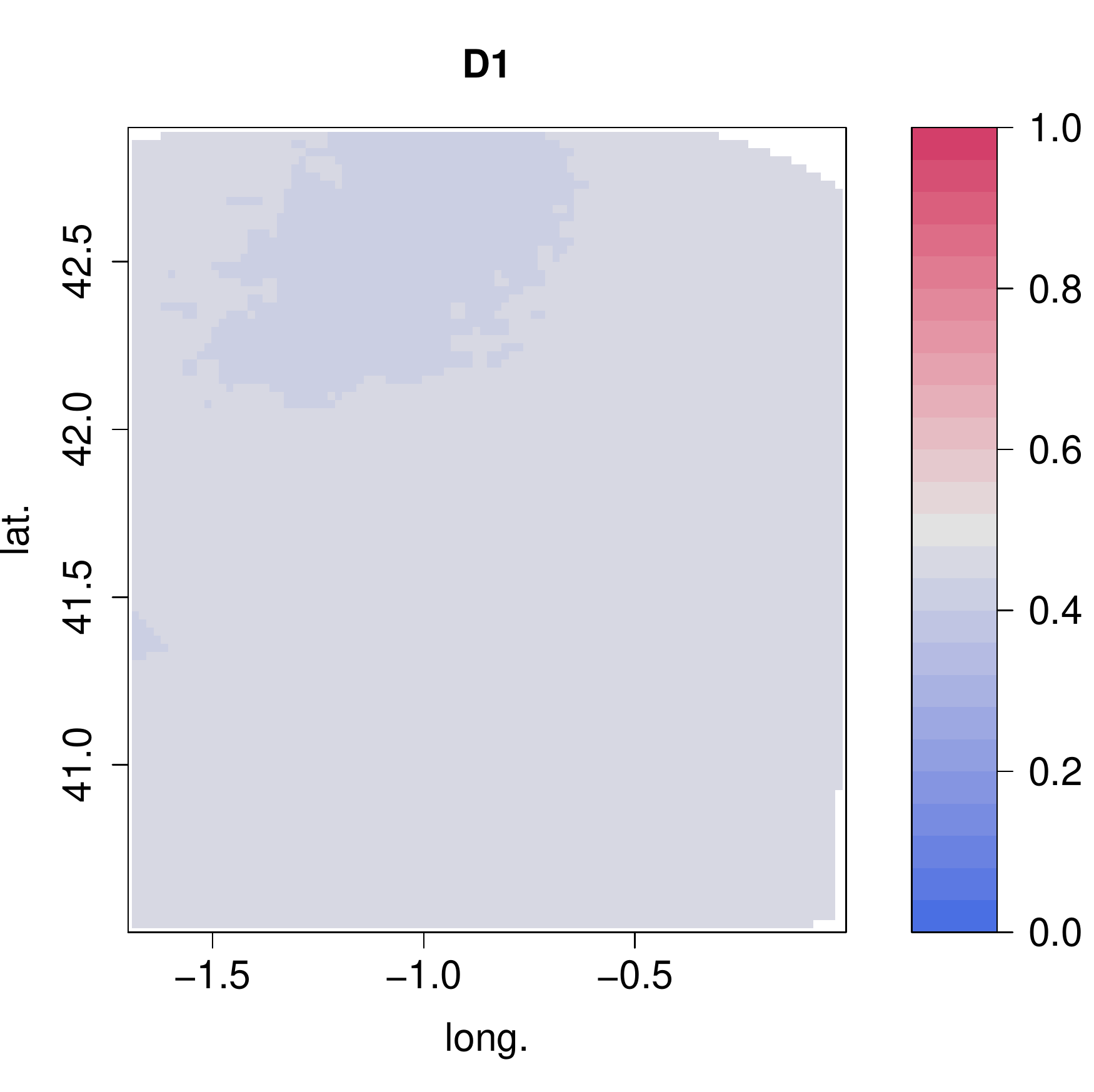}
		\includegraphics[width=0.31\textwidth]{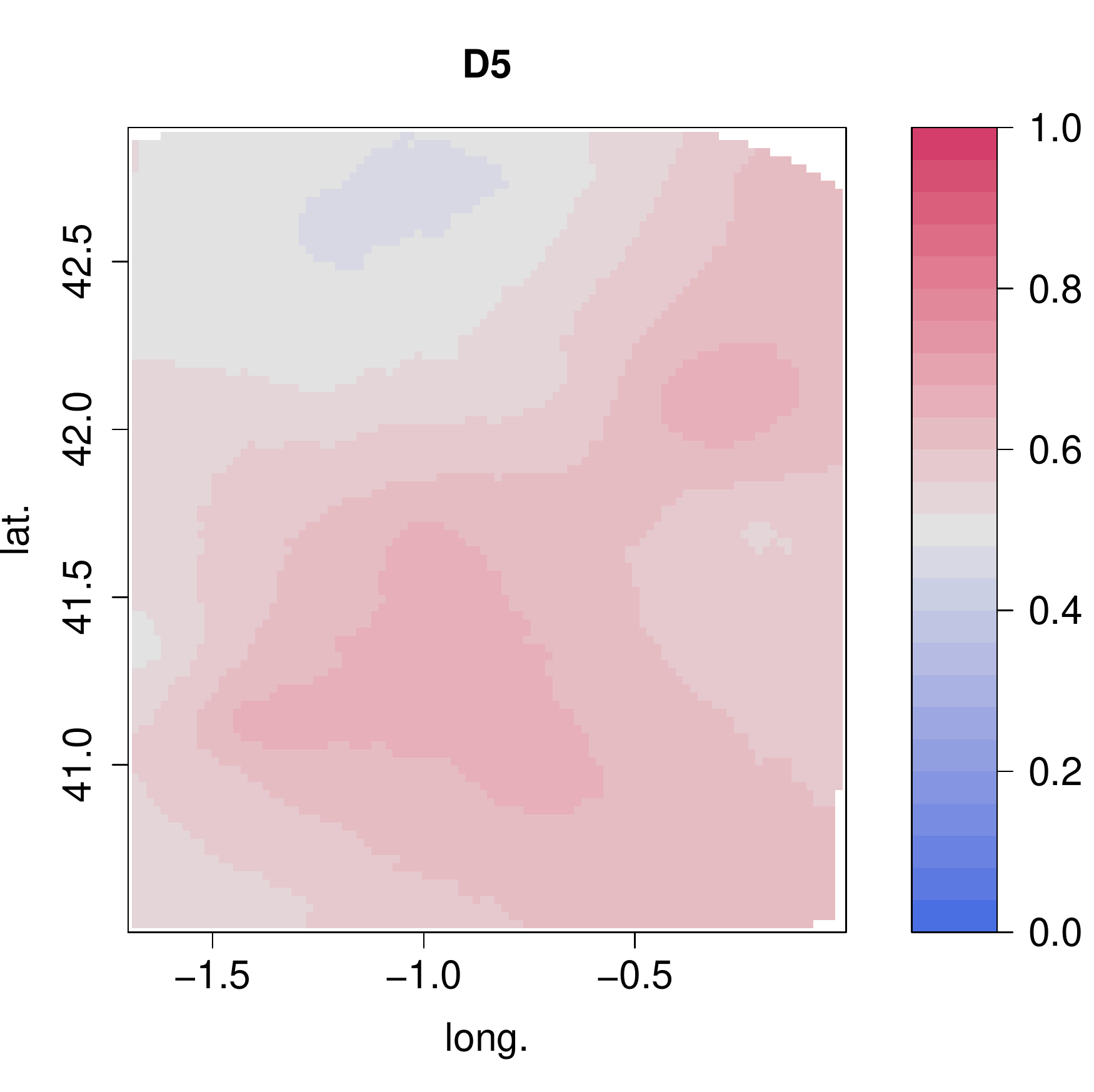}
		\includegraphics[width=0.31\textwidth]{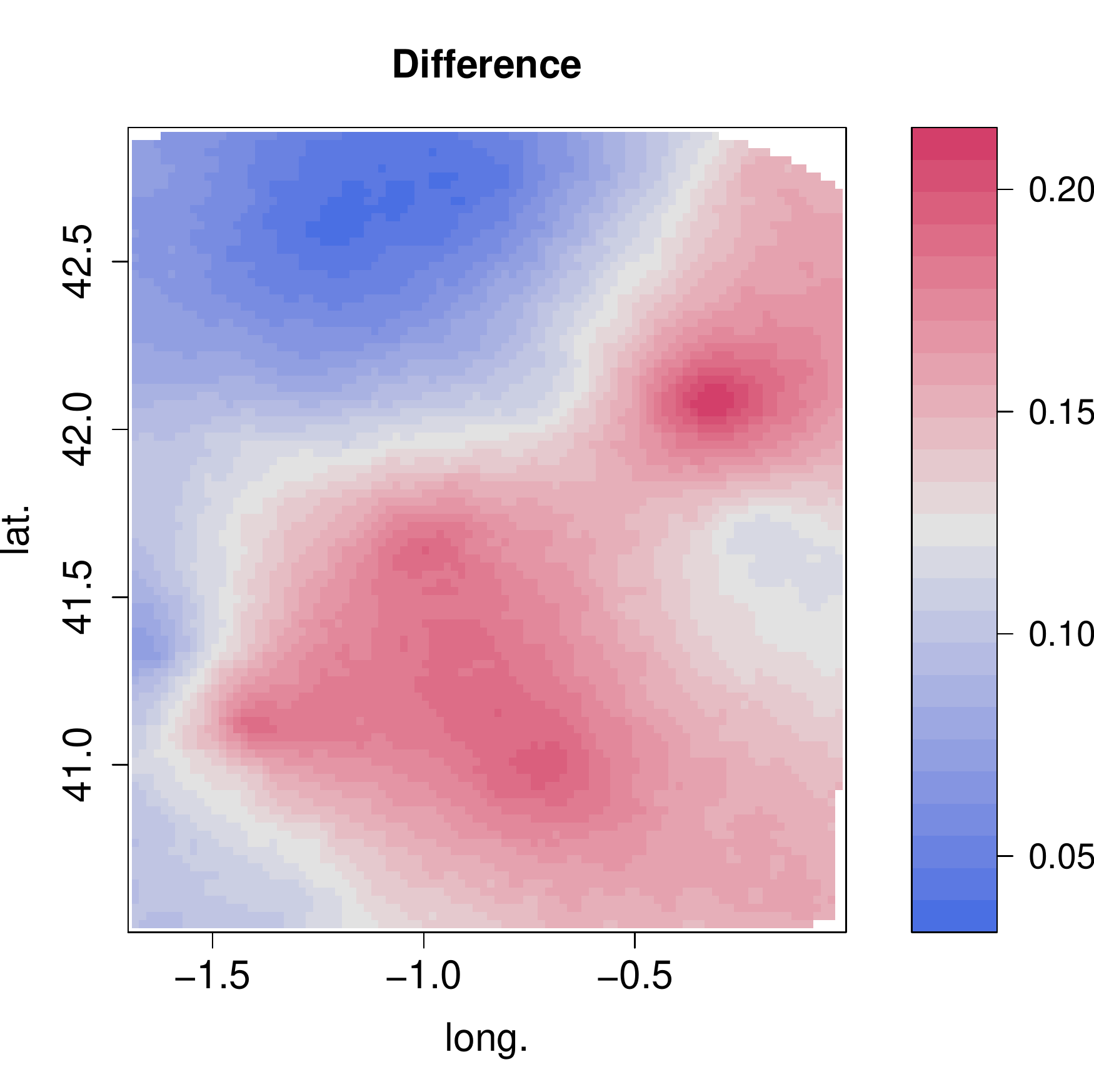}
		\includegraphics[width=0.31\textwidth]{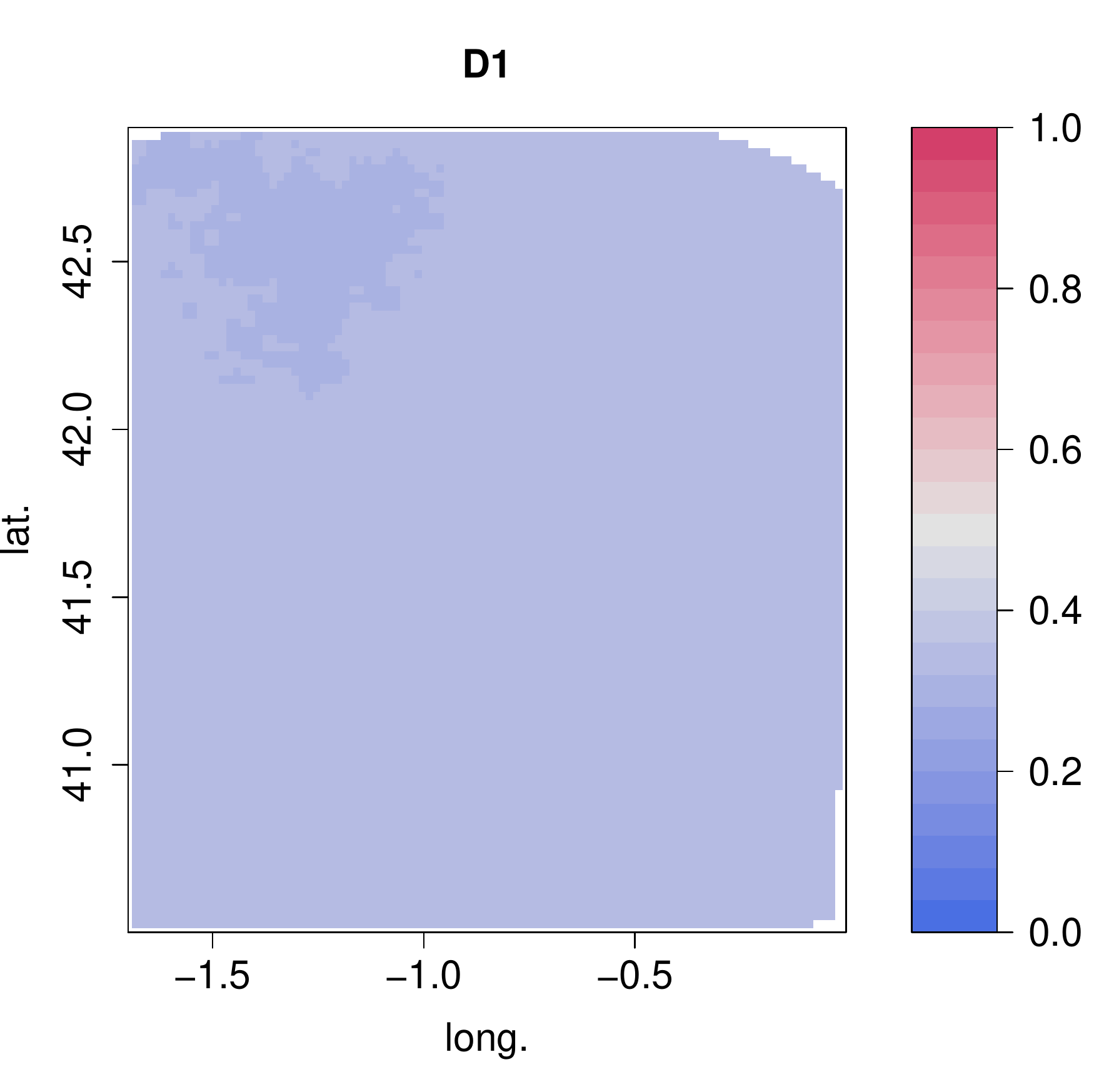}
		\includegraphics[width=0.31\textwidth]{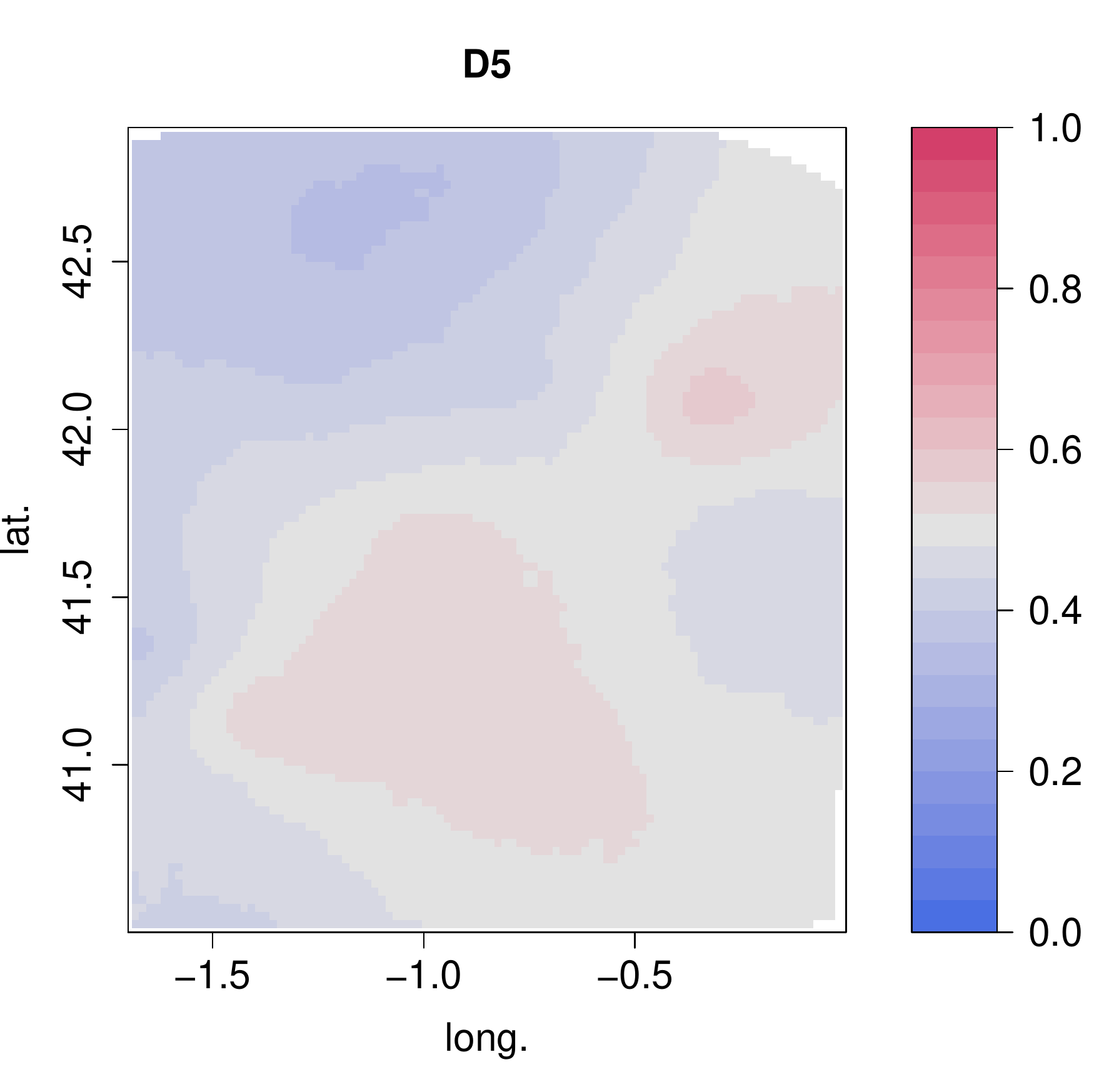}
		\includegraphics[width=0.31\textwidth]{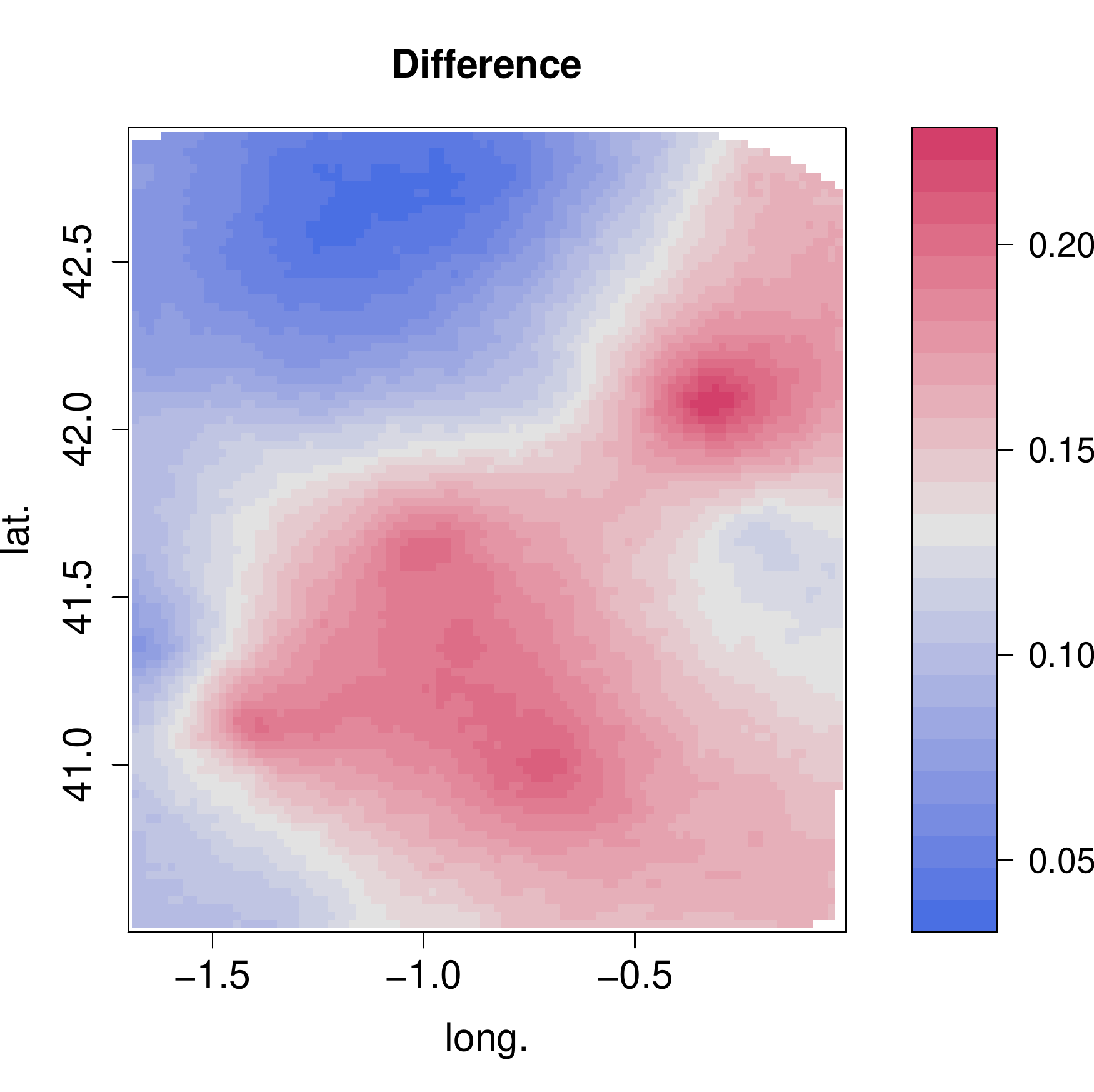}
	\end{center}
	\caption{Average  in D-JJA  of   probabilities  of  events  $\{Y_{t,\ell}(\bs)-\tilde \mu(\bs)>0\}$ (first row), and   $\{Y_{t,\ell}(\bs) -\tilde \mu(\bs)>0; 2 \}$ (second row)  in $D1$ and $D5$, and difference between them.}\label{F4}
\end{figure}

\medskip 

\textit{\textbf{Surface of  probabilities}}

The daily  posterior probabilities of  the previous events are computed using expression \eqref{EqP}, and averaged  over D-JJA using expression \eqref{Eq0b}. 
In $D1$, the probabilities of exceeding the  reference mean, $\{Y_{t,\ell}(\bs)-\tilde \mu(\bs)>0\}$,  vary from 0.42 to 0.47. However,  a clear evidence of global warming is observed in $D5$, 
since the probabilities all over the region are higher, attaining values close to  0.7  in 
the SW (the area  from Zaragoza, next to the Ebro river, to  Daroca and Cueva Foradada,  with a higher elevation) and also in the NE (Pyrenees area that contains Sallent and  Panticosa, the locations with highest elevation in the observed dataset). That means  that, in those areas, the  reference mean corresponds to the 30th percentile of the temperature distribution during $D5$. There is also evidence of changes in temperature persistence,  since the analysis  of the events   $\{Y_{t,\ell}(\bs) -\tilde \mu(\bs)>0; 2\}$  shows that the risk of   positive increments over the  reference mean during two consecutive days is around 0.3 in $D1$,  and in $D5$ it varies from  the same value 0.3 in the NW, to 0.6.

\medskip 

\textit{\textbf{Extents}}

\begin{table}[tb]
	\begin{center}
		\caption{Posterior mean  of the average extent in D-JJA for   increments  of daily temperature over the reference mean,   for different  values $c$  and persistence  in  decades $D1$ (1966-1975) and $D5$ (2006-2015); last row shows the mean of the extents for increments  of average temperature.
			\label{TDecades}}
		\begin{tabular}{l|cc|cc|cc}
			\multicolumn{1}{l|}{$c$} & \multicolumn{2}{c|}{$0^\circ$C} & \multicolumn{2}{c|}{$1^\circ$C}  & \multicolumn{2}{c}{$2^\circ$C}  	\\
			Decade & $D1$ & $D5$ & $D1$ & $D5$  & $D1$ & $D5$	\\
			\noalign{\smallskip}\hline\noalign{\smallskip}			
			$\{Y_{t,\ell}(\bs)-\tilde \mu(\bs)>c\}$			   & 0.45  & 0.58 & 0.37 & 0.50 & 0.29 & 0.41  \\
	$\{Y_{t,\ell}(\bs)-\tilde \mu(\bs)>c; 2\}$  & 0.34  & 0.47 & 0.26 & 0.38 & 0.19 & 0.30   \\
			$\{Y_{t,\ell}(\bs)-\tilde \mu(\bs)>c; 3\}$			    & 0.26  & 0.39 & 0.19 & 0.31 & 0.13 & 0.23  \\ 
			\noalign{\smallskip}\hline\noalign{\smallskip}
			$\{\bar Y_{D}(\bs)-\tilde \mu(\bs)>c\}$		 & 0.26  & 0.78 & 0.05 & 0.47 & 0.01 & 0.17  \\
		\end{tabular}
	\end{center}
\end{table}

Here, expression  \eqref{Eq2} is used to compute extents over the entire region associated with the foregoing events. The  average extent for events  based on daily temperature are computed  employing different periods of time.
First, we compute  yearly averages $		\overline{Ext}\left( A_t(\mathcal{B}) \right)=\frac{1}{92}\sum_{\ell\in JJA}  \widetilde{Ext}\left( A_{t,\ell}(\mathcal{B}) \right)$
to study the evolution across years of the events $ A_{t,\ell}(\bs) =\{Y_{t,\ell}(\bs)-\tilde \mu(\bs)>0\}$. Figure \ref{F6reg} (black line) shows the   posterior means of those  yearly averages, revealing a roughly linear increase  with  a trend  equal to $0.0035$ and 90\% credible interval (CI) $(0.0030, 0.0039)$; this means  an increase in extent per decade equal to $3.5$\%. A similar  evolution across years is expected in the  extent for events defined with  different increments and persistence; e.g., the linear trends  for events with increments higher than $c=1$ and  $2^\circ$C are equal to  $0.0035$ and $0.0033$, respectively.  As an aside, the trend  of  empirical extents, i.e., the proportion of observed stations exceeding their reference value, shown in Fig. \ref{FigExplor_Extents} is similar, $0.0037$, for temperatures over the reference mean.
However, an evident limitation of this empirical extent  is that  uncertainty of the empirical extents cannot be quantified. Moreover,  it is defined relative to only 18 stations as opposed to the fine grid of 4401 locations employed in our posterior predictive simulation.

Regarding the average extents over decades $\overline{Ext}\left( A(\mathcal{B}) \right)$, see   expression \eqref{Eq3},  Table \ref{TDecades}  summarizes their means   for events $\{Y_{t,\ell}(\bs)-\tilde \mu (\bs)>c\}$  with $c=0,1,2^\circ$C  in  $D1$ and $D5$, and for  the persistent events defined with 2 and 3  consecutive days. The variability of the  average extents is quite low, with 90\% CI of length around 0.06 in all the cases. This variability is much lower than the variability  across decades,  indicating a clear  increase  in the extent for all types of events; e.g., the   mean  and the 90\% CI of the average extent  for daily temperatures over $\tilde \mu(\bs)$ in  $D1$ and  $D5$ are respectively, $0.45 \ (0.42, 0.48)$ and  $0.58 \ (0.55, 0.61)$.  That increase  yields a similar extent for events  $\{Y_{t,\ell}(\bs)-\tilde \mu(\bs)>0\}$ in $D1$  and the extent for  events $\{Y_{t,\ell}(\bs)-\tilde \mu(\bs)>2\}$ in  $D5$, that is $0.41 \ (0.38, 0.45)$.  As a consequence of this warming, the average extent in $D5$ with $c=1^{\circ}$C is higher than the average in $D1$ with $c=0^{\circ}$C. The increase is also  observed in the extent for   persistent events based on three days,  especially  in increments higher than  $c=2^{\circ}$C,  where the mean  of the average extent in $D5$ shows a relative increase with respect to $D1$, higher than 75\%, from 0.13 o 0.23.

\subsubsection{Analysis of  increments of average temperature  over $\tilde\mu(\bs)$}

This section summarizes  the analysis of  events  based on the average temperature in D-JJA, $\{\bar Y_{D}(\bs) -\tilde \mu(\bs)>c\}$ for decades $D1$ and $D5$ and values $c=0,1$ and $2^{\circ}$C.

\medskip 

\textit{\textbf{Surface of  probabilities}}

\begin{figure}[t]
	\begin{center}
		\includegraphics[width=0.31\textwidth]{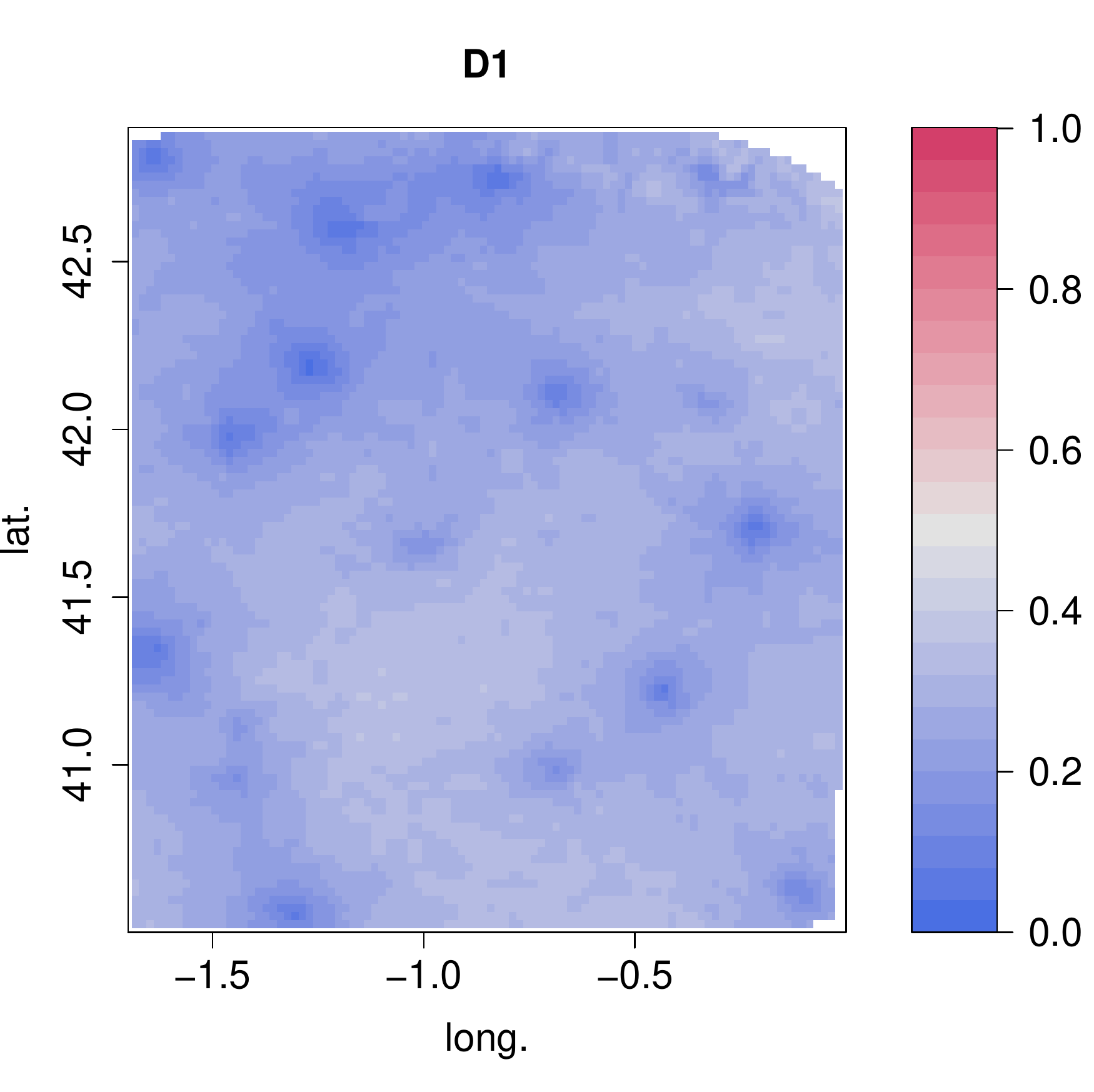}
		\includegraphics[width=0.31\textwidth]{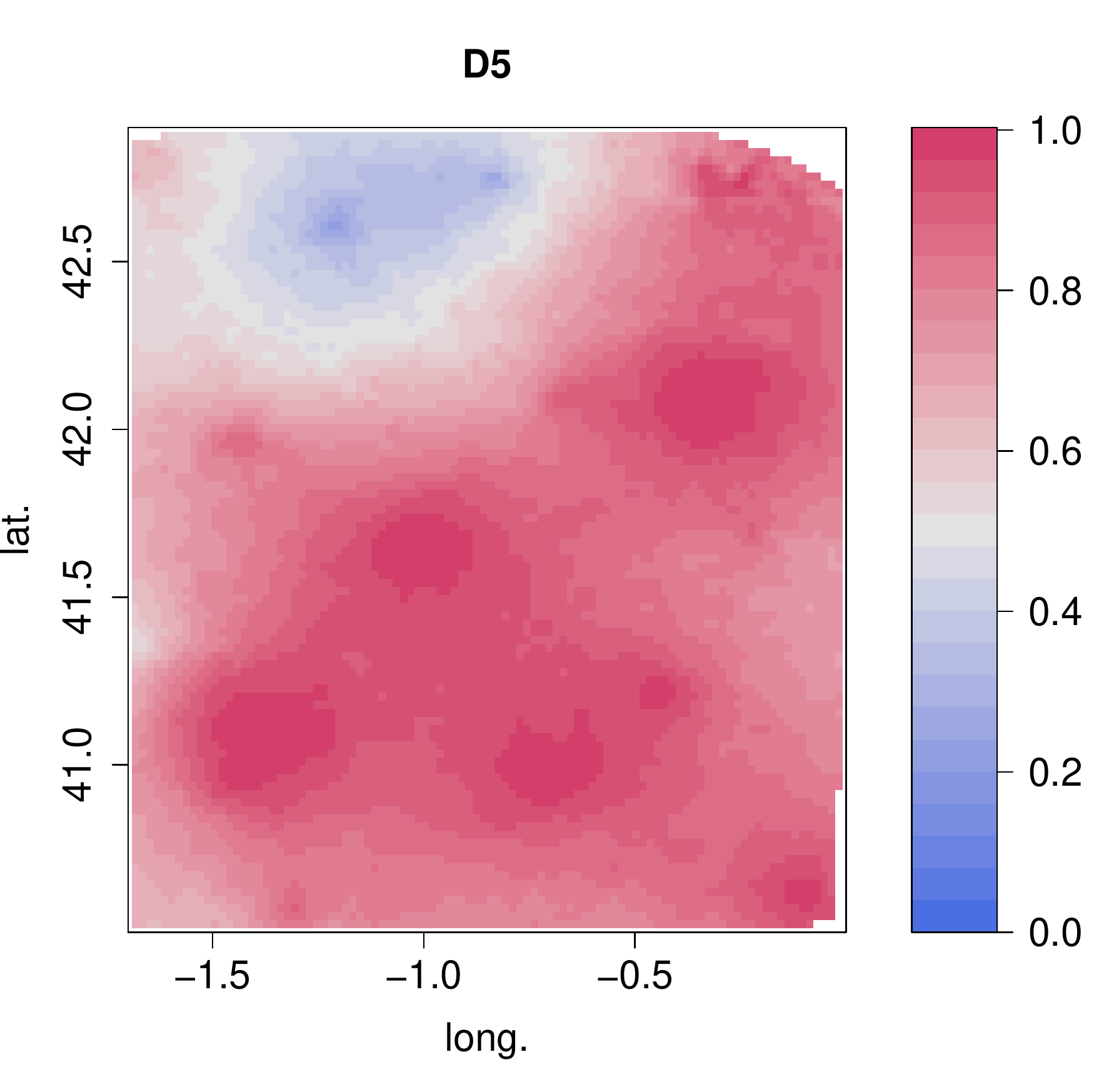}
			\includegraphics[width=0.31\textwidth]{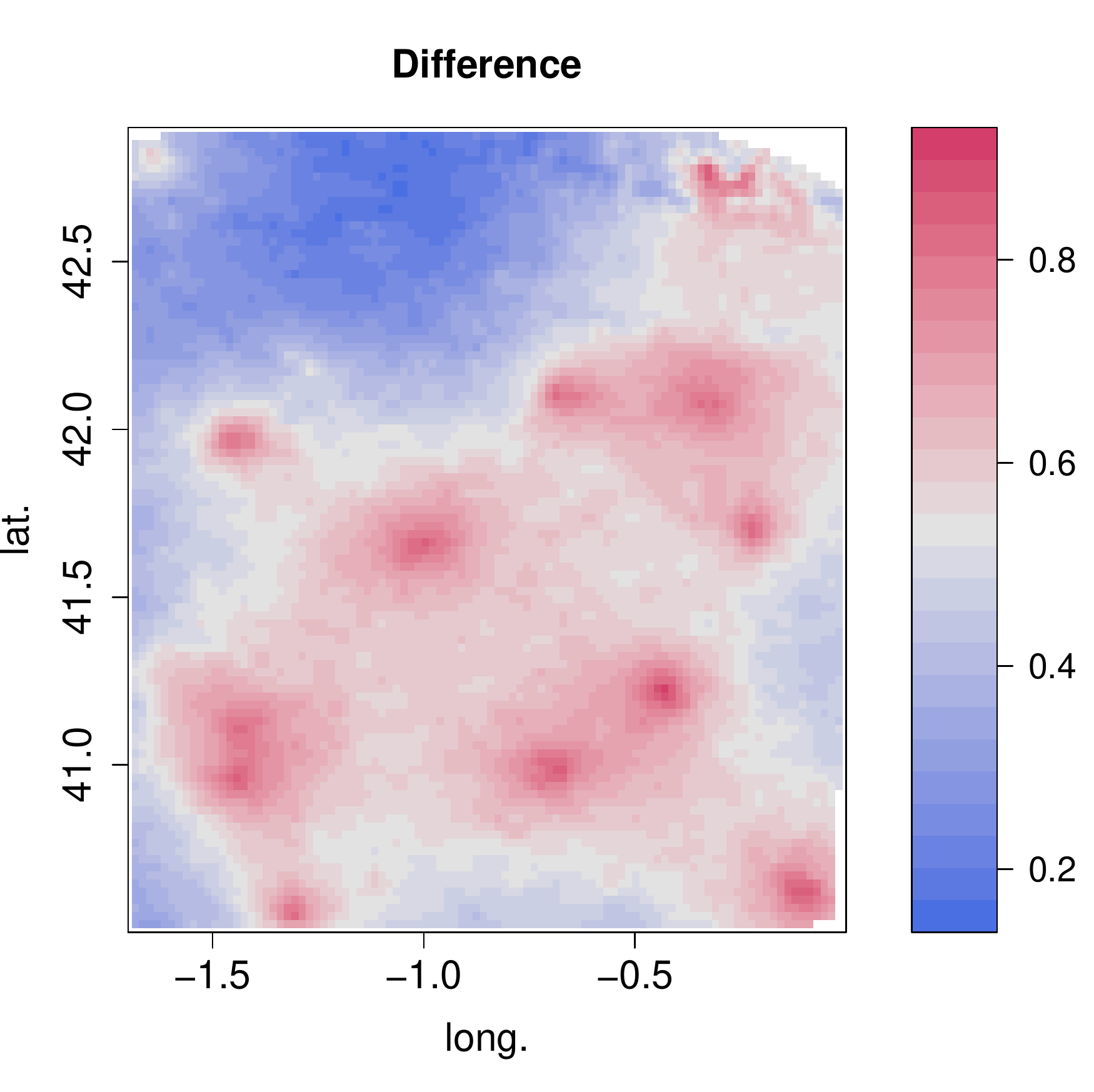}	
		\includegraphics[width=0.31\textwidth]{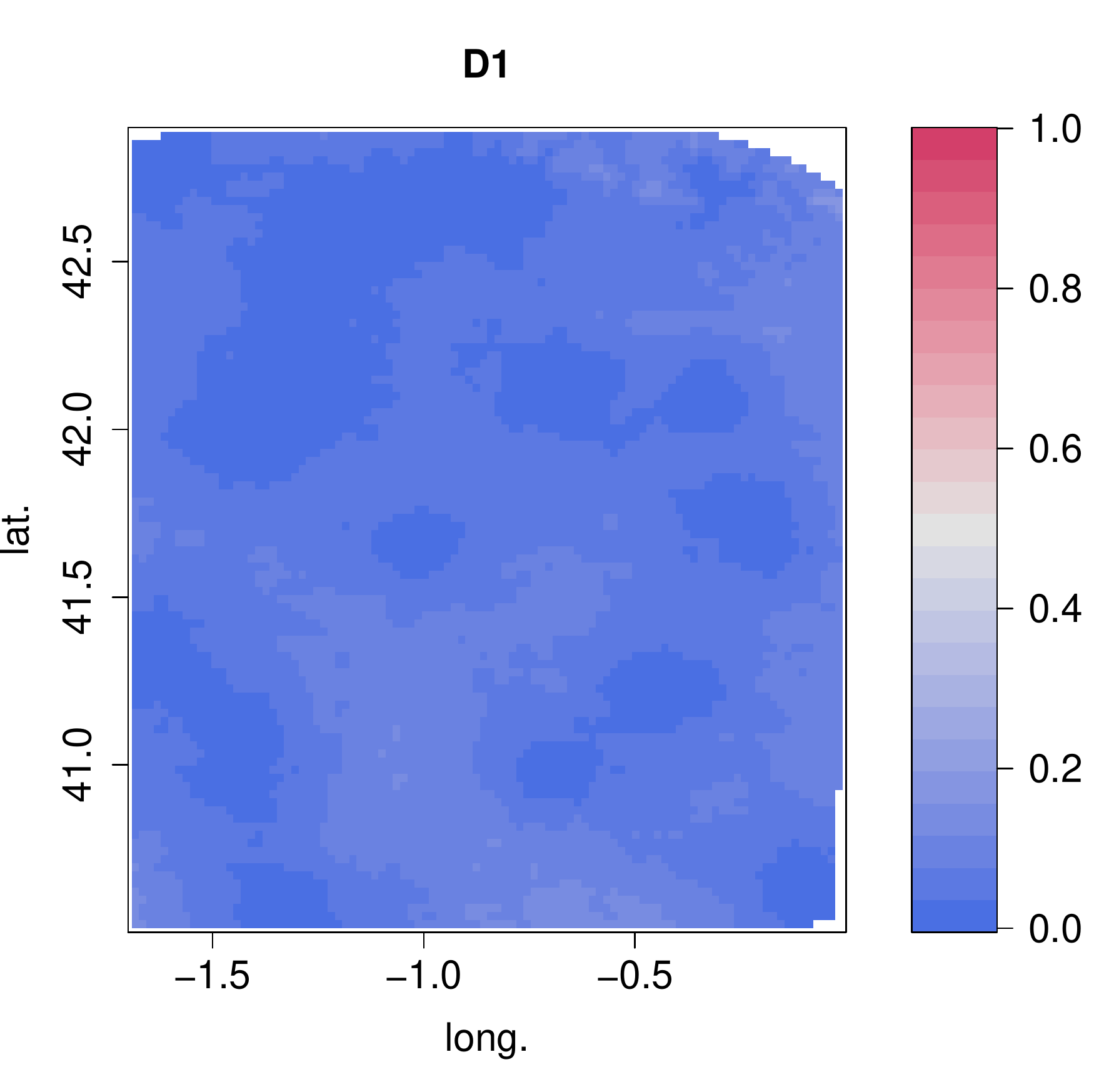}
		\includegraphics[width=0.31\textwidth]{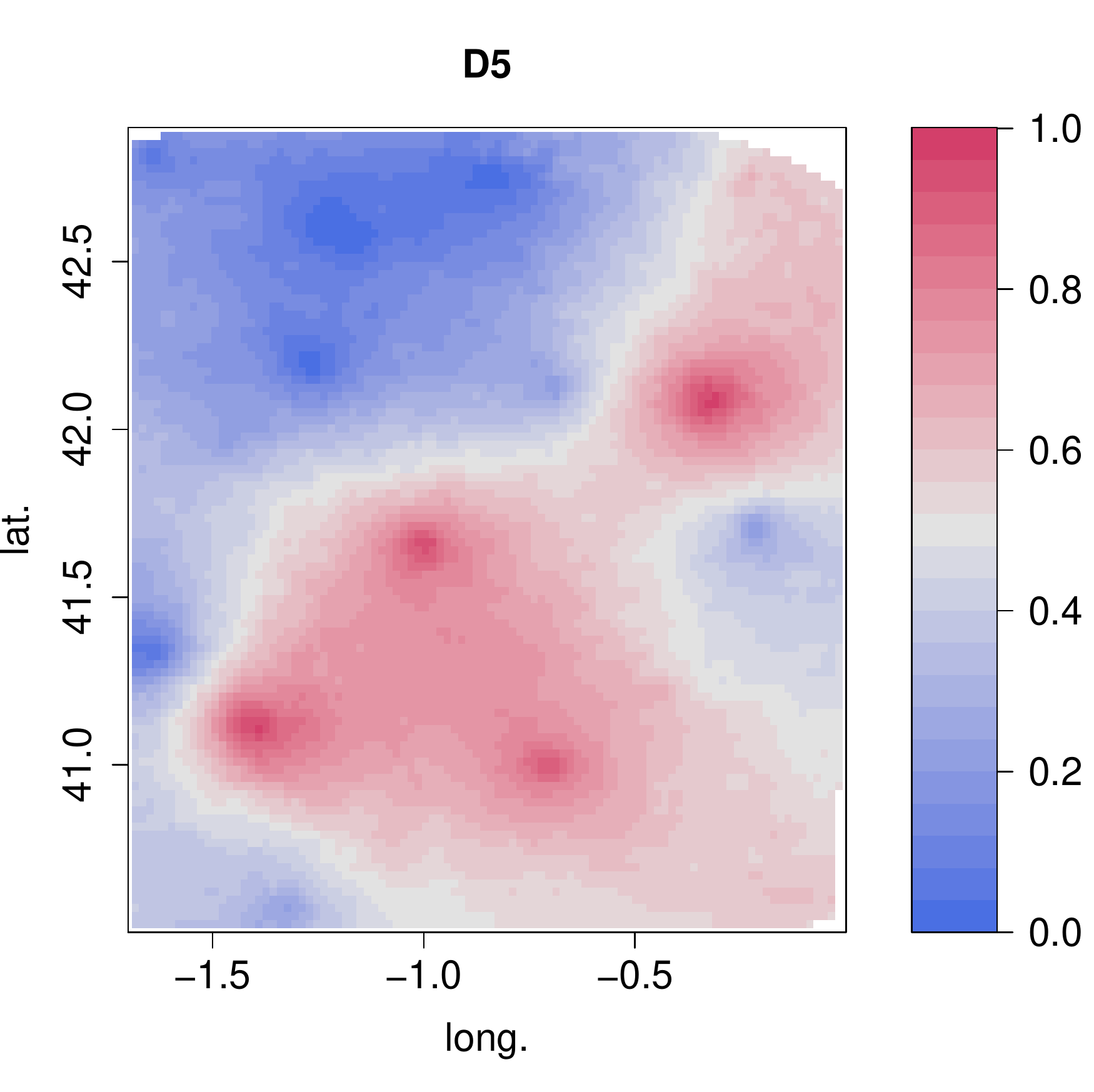}
		\includegraphics[width=0.31\textwidth]{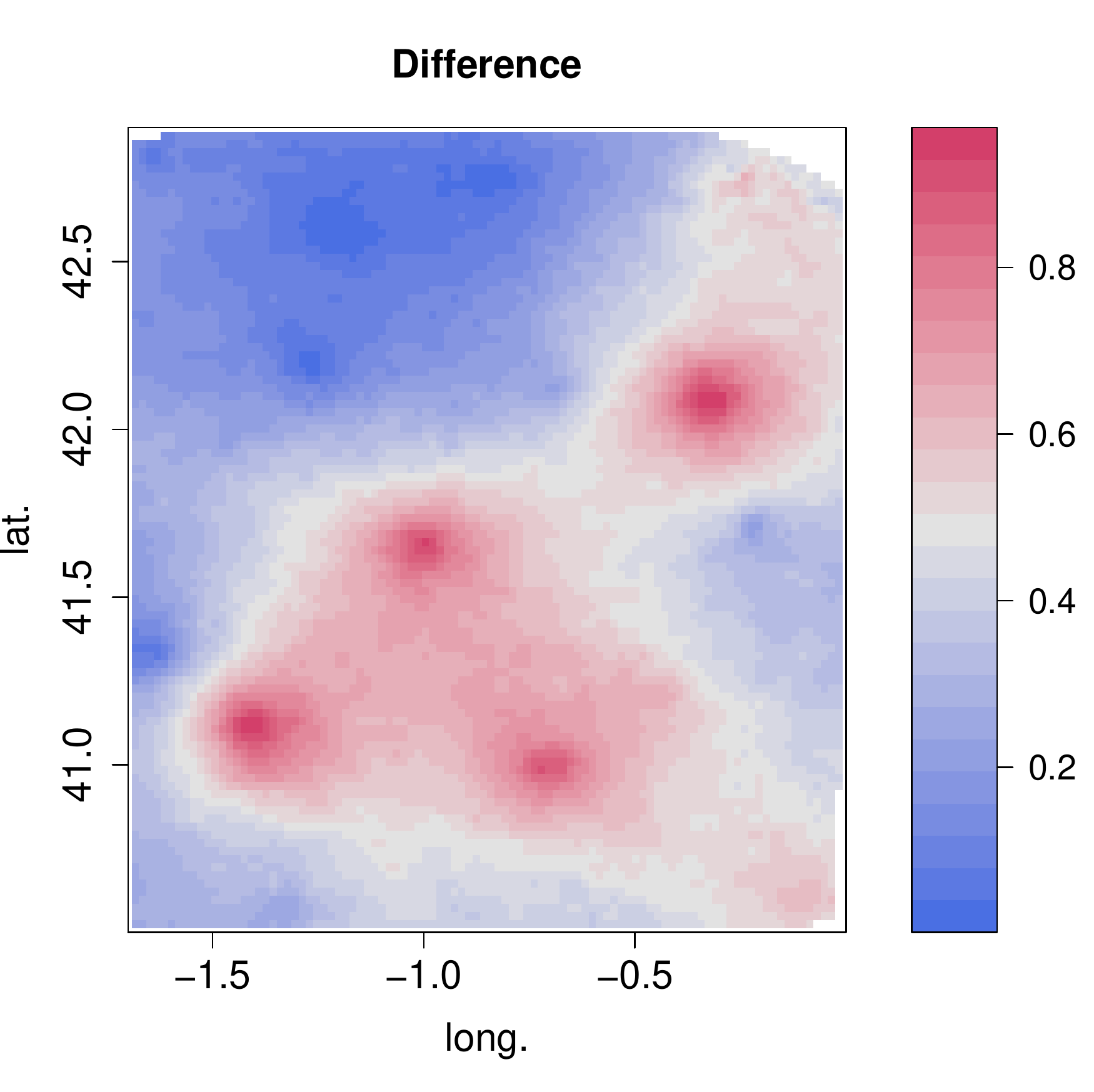}
	\end{center}
	\caption{
	Probabilities of events  $\{\bar Y_{D}(\bs) -\tilde \mu(\bs)>0\}$ (first row),  and $\{\bar Y_{D}(\bs) -\tilde \mu(\bs)>1\}$ (second row)  in decades $D1$  and $D5$  and differences between them.}\label{F3} 
\end{figure}

In $D1$, the  risk of  average temperature higher than $\tilde \mu(\bs)$  varies slightly throughout the region, from 0.03 to  0.4. In $D5$,    this risk  is much  higher (from 0.7 to virtually 1)  in all the region except in the NW, the area closer to  the Cantabrian Sea, see Fig. \ref{F3}.   The pattern of the increase in the risk of this event is different from most of the other events  where  the areas with highest  risk of suffering the effects of climate change are   the center of the valley and  the NE  areas.
Regarding the risk of   increments of the average temperature over $\tilde \mu(\bs)$ being higher than $1^{\circ}$C, in $D1$, it is quite homogeneous throughout  the region: lower than 0.08 in 75\% of the region and always lower than 0.2. However,    although  the risk in $D5$ has increase  all over the region, there are relevant  differences depending on the area: it varies  from   values lower than 0.2 in the NW up to more than $0.7$ in the central part of the valley.

\medskip 

\textit{\textbf{Extents}}

\begin{figure}[t]
	\begin{center}
		\includegraphics[width=0.85\textwidth]{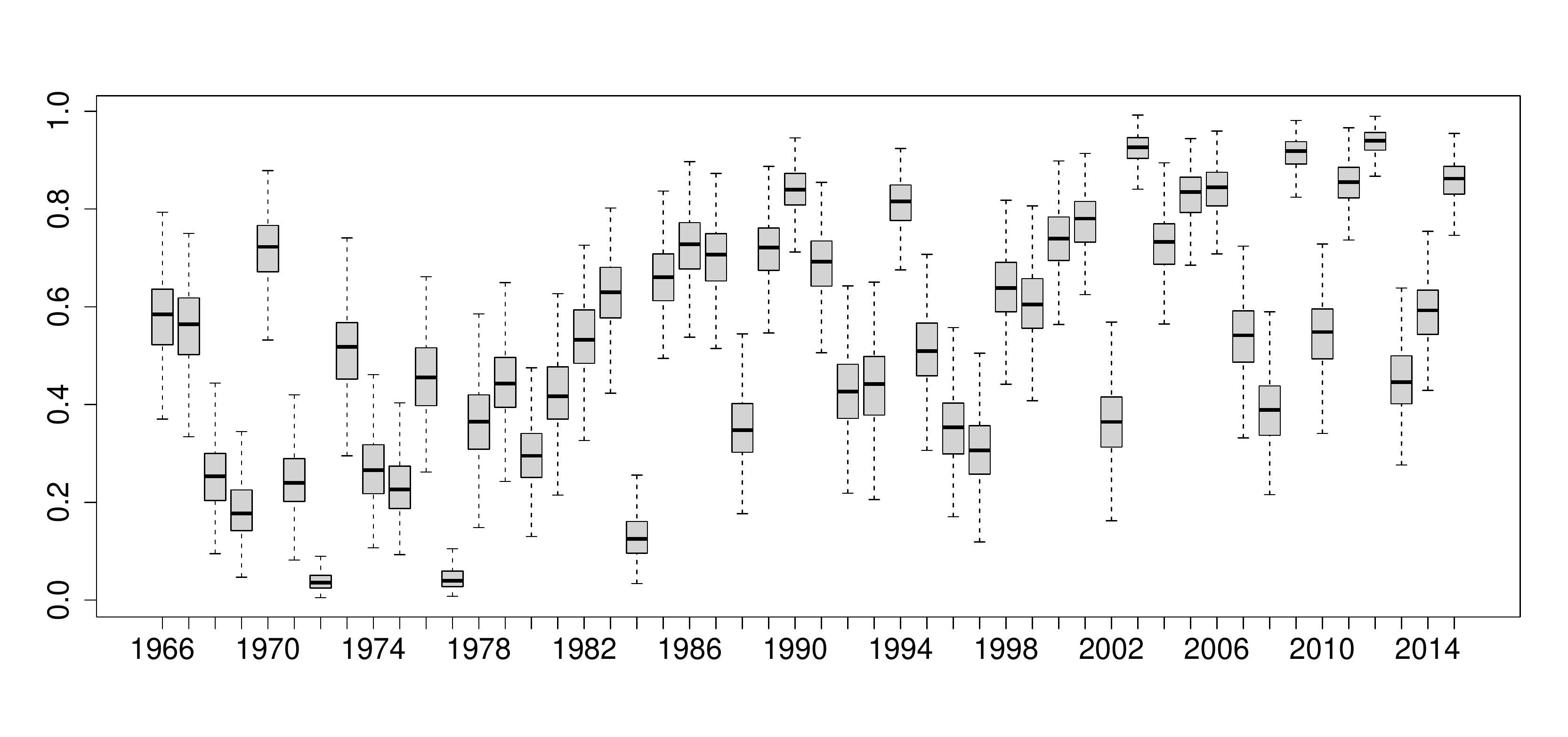}	
	\end{center}		
	\caption{		Boxplots of  the distribution of the extent  for events based on yearly average temperatures $\{\bar Y_{t}(\bs)-\tilde \mu(\bs)>0\}$,  versus year. \label{F6}}
\end{figure}

First, to characterize  the evolution over time, we compute  the extent  of positive increments of the average temperature in JJA in each year $\{\bar Y_{t}(\bs)-\tilde \mu(\bs)>0\}$.    Figure \ref{F6}   shows the  boxplots of the  posterior distribution of those yearly  extents. The  increasing trend  of the  extent  is clear, demonstrating that the variability between years is higher than the variability within year. The slope of these extents is $0.0088$, more than double the slope of the extents based on daily temperatures.
In addition to the increasing trend, this plot permits us to identify  years which were colder with respect to  the  trend, and with a lower variability,  e.g., 1972, 1977, 1984, or hotter as year 2003 \citep{Garcia15}. In the last decade, two different types of behaviors are observed,  the distribution of  the extent in some years is  quite high, centered around 0.9, while others centered around  0.5.

\begin{figure}[t]
	\begin{center}
		\includegraphics[width=0.45\textwidth]{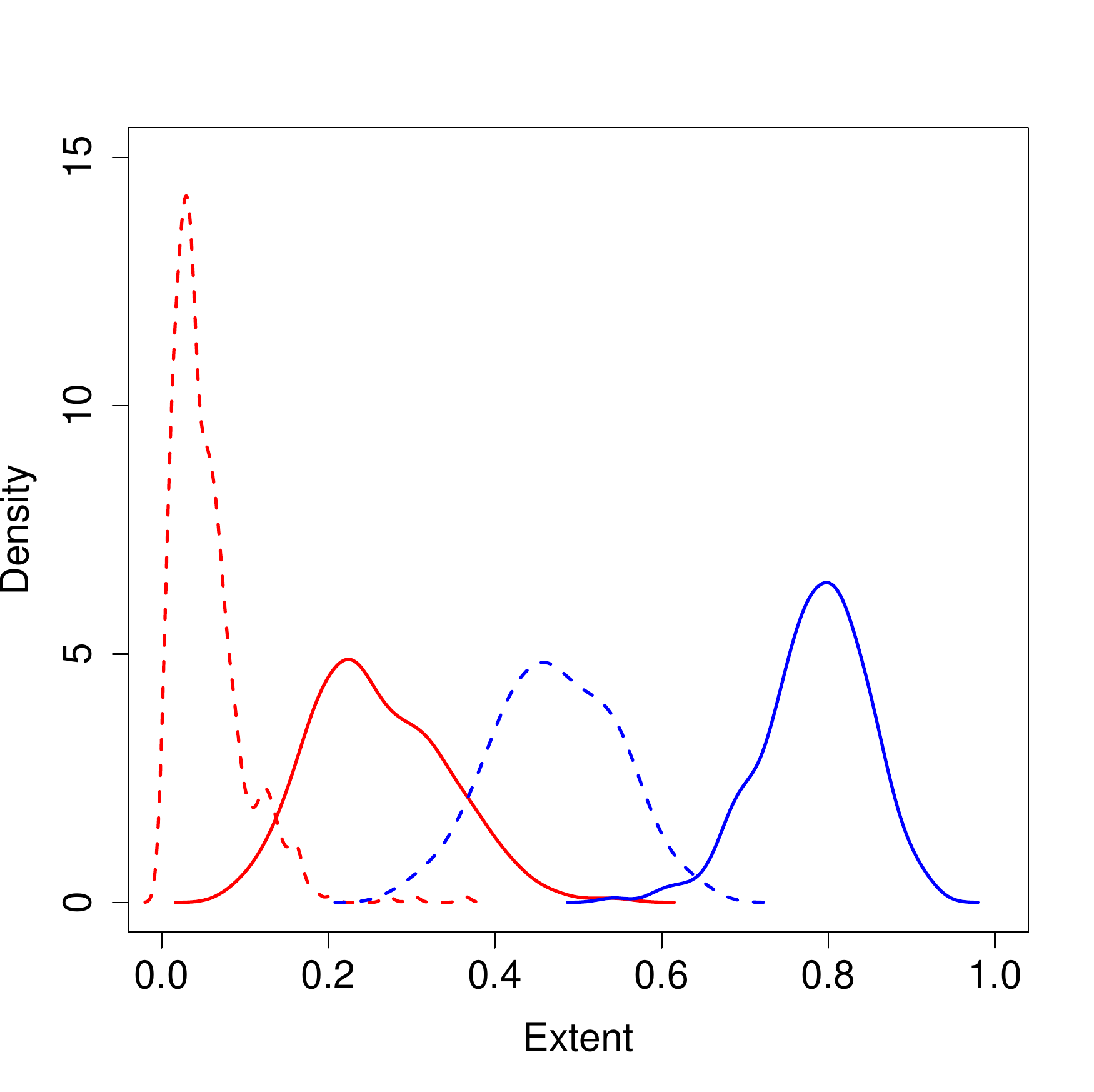}	
	\end{center}		
	\caption{Posterior density of the extent for events  $\{ \bar Y_{D}-\tilde \mu(\bs)>c\}$ with $c=0$ (solid line)  and $c=1$ (dotted line) in  $D1$ (red) and $D5$ (blue). 
		\label{F5Ext}}
\end{figure}

We also analyze  the extent  for increments of decadal averages $\{\bar Y_{D}(\bs)-\tilde \mu(\bs)>c\}$; the last row in Table \ref{TDecades} summarizes the posterior mean of  those extents with $c=0,1,2^\circ$C  and Fig. \ref{F5Ext} compares their posterior densities   in $D1$ and $D5$ for $c=0$ and 1.    The ratio of the mean extents in $D5$ and $D1$  increases with $c$:  it is equal to   3 for $c=0$,  9.4 for $c=1$, and 17 for $c=2^\circ$C.  The variability of the  posterior distribution of these extents is not large so that the probability of the extent for events $\{\bar Y_{D}(\bs)-\tilde \mu(\bs) >0\}$ being higher in $D5$ than in $D1$ is virtually 1 for  the three $c$ values. A consequence of this increase is that the  mean of the extent   for increments higher than 0 in $D1$ is roughly one third  its counterpart in $D5$, and almost  half  the extent    of increments higher than $1^\circ$C  in $D5$.   

It is noteworthy  that  the analysis of both probabilities and extents shows that  consequences of global warming are stronger  in average temperatures than in daily temperatures.

\subsubsection{Analysis of temperature increments between decades}

\label{S43}

This section summarizes  the  analysis of events that  quantify the global warming in terms of the   increments of average temperatures, $\{\bar Y_{D5}(\bs)-\bar Y_{D1}(\bs)>c\}$ for values $c=0,1$ and $2^{\circ}$C.

\medskip 

\textit{\textbf{Surface of  probabilities}}

\begin{figure}[t]
	\begin{center}
		\includegraphics[width=0.32\textwidth]{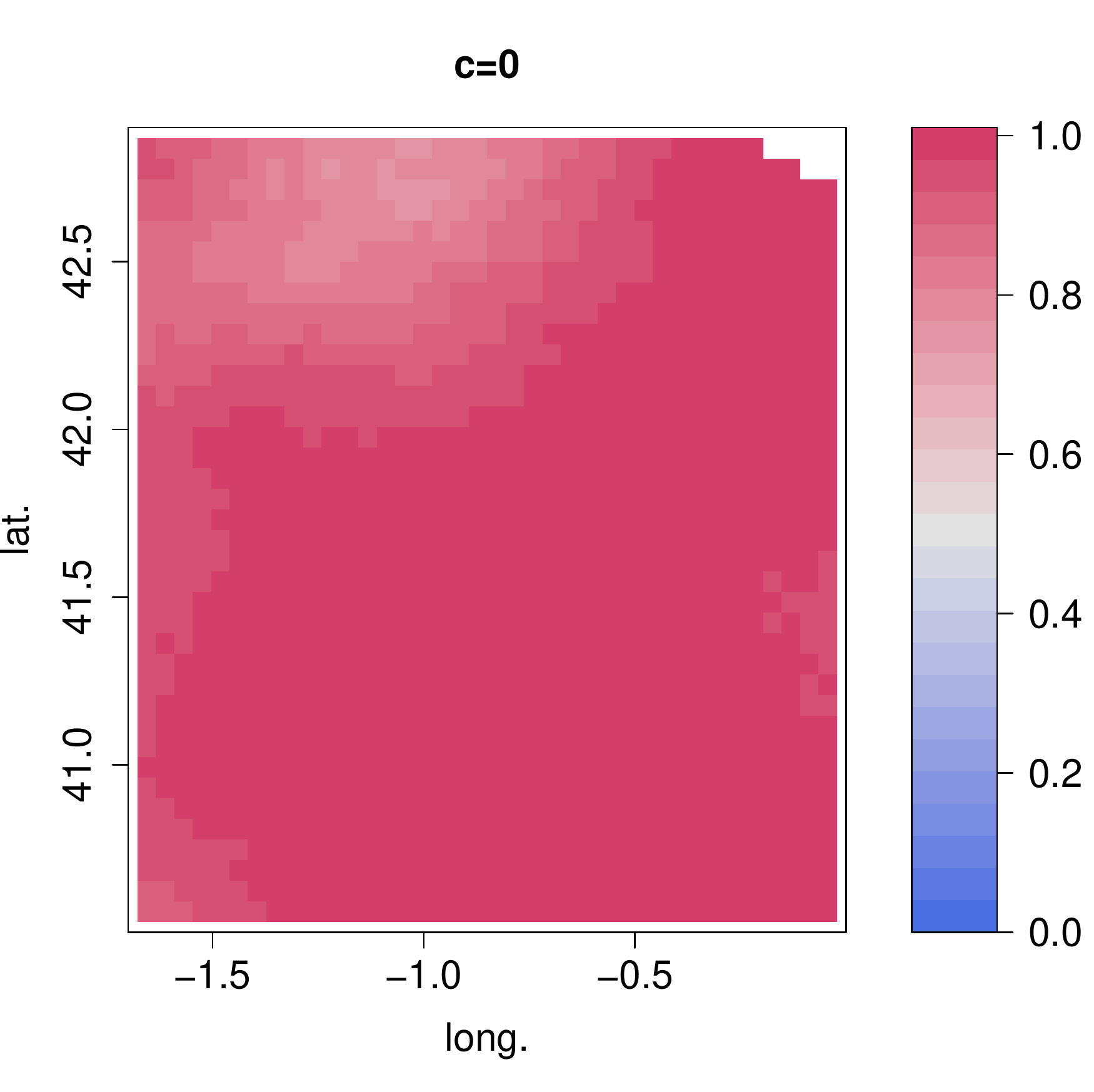}
		\includegraphics[width=0.32\textwidth]{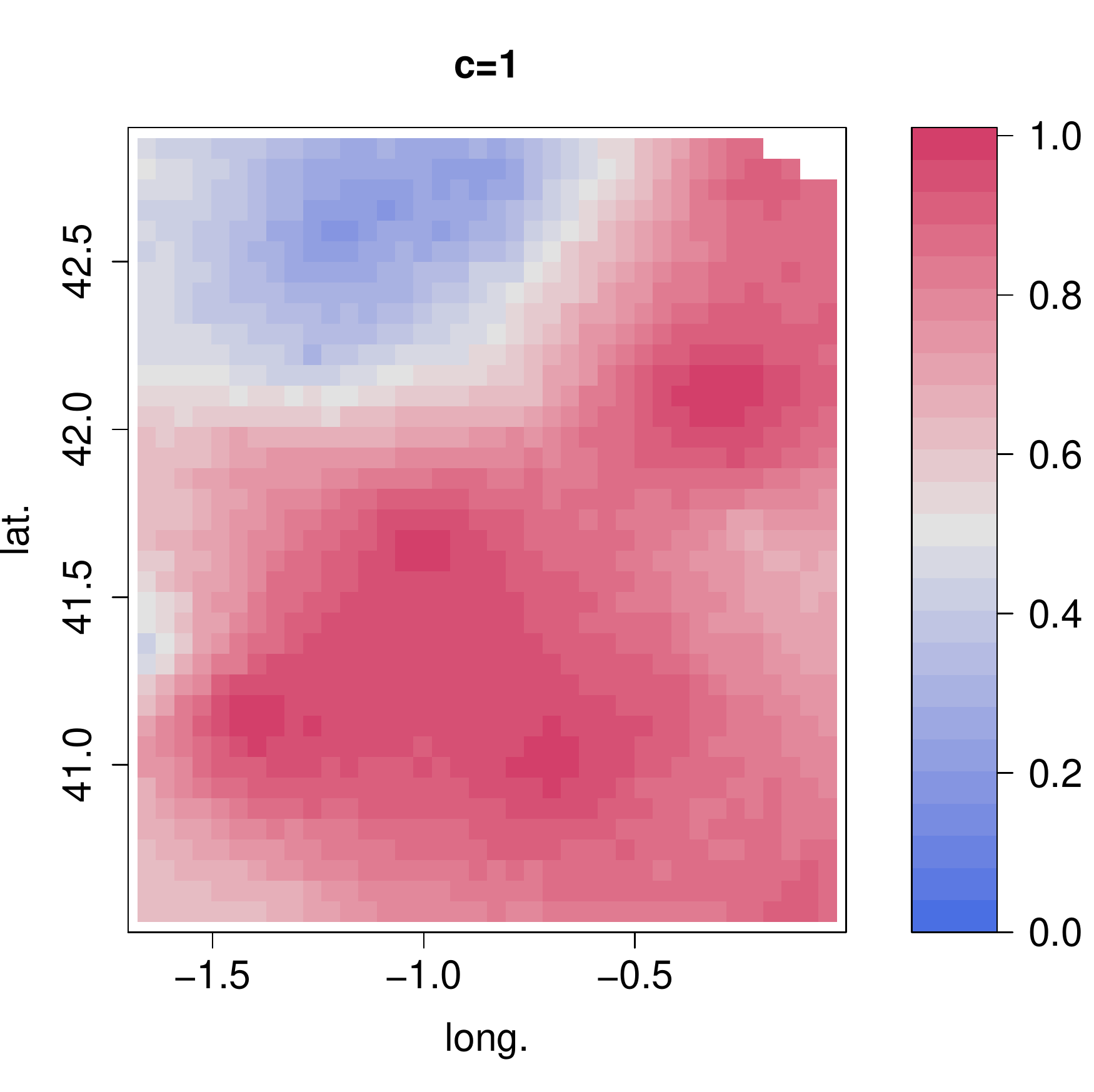}
		\includegraphics[width=0.32\textwidth]{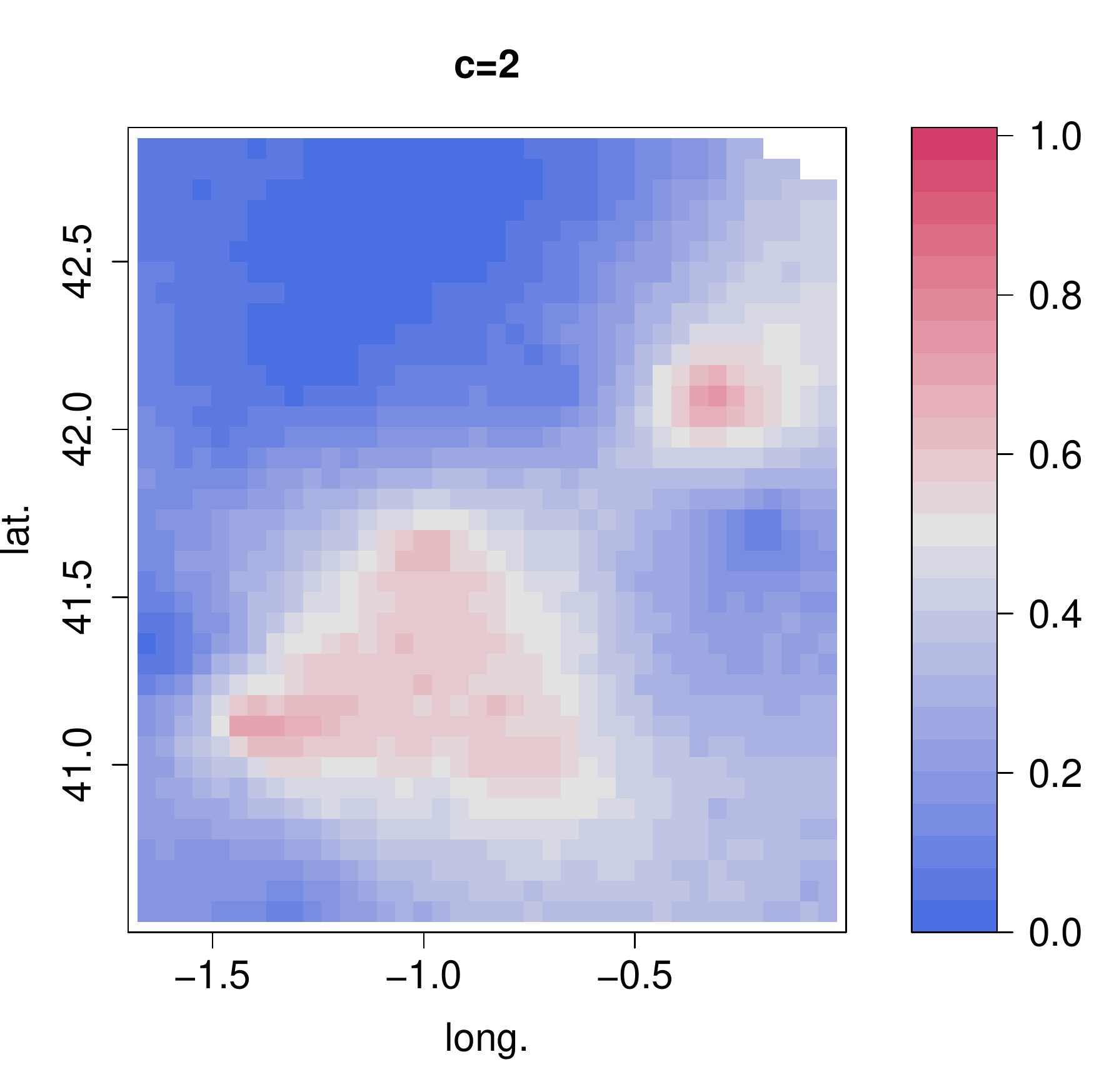}				
	\end{center}
	\caption{Posterior probabilities of  increments of average temperatures $\{\bar Y_{D5}(\bs)-\bar Y_{D1}(\bs)>c\}$ for $c=0,1,2^\circ$C. \label{F3b} }
\end{figure}

According to Fig. \ref{F3b} the risk of a positive increment of average temperatures between $D1$ and $D5$ is virtually  1  all over the region, except in the NW where it takes values around 0.7. However,  for other $c$ values, the spatial variability is higher.  The risk of  an increment higher than $1^\circ$C,  is  close to 1 in  some areas and   higher than 0.6 except in the NW where  it is roughly 0.25. The risk of increments higher than  $2^\circ$C is lower than 0.4 in most of the region except some  small areas in  the center of the valley and the NE, where it attains 0.7. Comparing these results with the analysis of daily increments in  the Supplement, Section S4.1, we note that the risk of  an increment  between $D1$ and $D5$  higher than $c$ is  much higher for  average temperatures  than for  daily temperatures.

\medskip 

\textit{\textbf{Extents}}

Table 	\ref{TINcReg2} summarizes the   means  and the $90\%$ CI of the extents   for increments between average temperature  in $D1$ and $D5$.
	The  CI of the extents show that  between 90 and 98\% of the area under study has suffered a positive increment of the average temperatures, from 58 to 80\% an increment  higher than $1^\circ$C,   and from 15 to 36\% higher than $2^\circ$C.

\begin{table}[t]
	\begin{center}
		\caption{Posterior mean  and CI of  the extent for events $\{\bar Y_{D5}(\bs)-\bar Y_{D1}(\bs)>c\}$ for   different regions $\mathcal{B}$ and values $c$.}
		\label{TINcReg2}
		\begin{tabular}{l|ccc}
			$c$	& $0^\circ$C & $1^\circ$C & $2^\circ$C \\ \hline
			$\mathcal{D} $	& 0.95  (0.90, 0.98) & 0.69  (0.58, 0.80) & 0.25  (0.15, 0.36)  \\ 
			$\mathcal{V}$	&  0.98   (0.95, 1.00) & 0.80   (0.68, 0.89) & 0.32   (0.18, 0.47)  \\	
			$\mathcal{P}$	&  0.93   (0.85, 0.98) & 0.63   (0.45, 0.79) & 0.19   (0.07, 0.36)  \\
		\end{tabular}
	\end{center}
\end{table}

\subsection{Comparison of the evolution in areas with different climates}

\label{S44}

The region considered in this analysis includes areas with very different climates,  see Fig. \ref{Fig_relief_climate}. Here, we analyze whether the  consequences of global warming  are the same over the entire region or whether we can identify  different patterns of evolution over time. This type of study is not possible using the observed database, since the number  of  available stations  in some areas is sparse. The  use  of the output from the statistical model enables that type of  comparison.  More precisely,  in this secction, we use the approach described in  Section \ref{S3}   to compute the extent  for different events in  two regions with different climates, and to compare the  effects of global warming in  those areas.

We consider  two  important regions in the study area, which according to the K\"{o}ppen's climate classification  have very different characteristics. Region  $\mathcal{V}$ (valley)  is  the area between parallels 41N and 42N with the semiarid  Bsk climate. It covers  the central Ebro valley and it has a mean elevation of 373 m. 
This area is the most populated in Aragón, and the most important farming areas in the region are located there.  Region    $\mathcal{P}$ (Pyrenees)  is a mountainous area in the Pyrenees,  over  parallel  42N,  with mountain climate Cfb  and some small areas with high mountain  climates Dfb and Dfc. The mean elevation is 1,427 m  but in  some  points the elevation is over  3,000 m. The last glaciers in Spain are located in this area.

\subsubsection{Average extents for increments of daily  temperatures over  $\tilde \mu(\bs)$}

Table \ref{TDecades2}  summarizes the mean of  the average extent in D-JJA for events $\{Y_{t,\ell}(\bs)-\tilde \mu(\bs)>c\}$,  for $c=0$ and $2^{\circ}$C  in decades $D1$ and $D5$ and  regions $\mathcal{V}$ and $\mathcal{P}$, and  for  the analogous events defined with $k=2$ and 3  consecutive days. 
	
	The increase in extent between decades $D1$ and $D5$ is observed in both regions,  but  it is clearly higher in $\mathcal{V}$. The mean of the average extent in $D1$ is quite similar in both regions. However,   clear differences appear in $D5$, specially for the mildest events with $c=0$: the mean of the percentage of area with daily temperatures higher than the reference mean  is 60\% in $\mathcal{V}$ and 54\% in  $\mathcal{P}$.  These differences become smaller in more exigent events; e.g.,   the mean of the percentage of area with increments over  $\mu(\bs)$ higher than $2^{\circ}$C during three consecutive days is  24\% in $\mathcal{V}$ and 20\% in  $\mathcal{P}$. However, in both regions the increase is clear since the counterpart in $D1$ is 13\%.

	\begin{table}[tb]
		\begin{center}
			\caption{Posterior mean  of the  average extent in D-JJA for increments of  daily  temperature over the reference mean higher than $c$ with different persistence   for  reference values $c=0,2^{\circ}$C,   in decades $D1$ (1966-1975) and $D5$ (2006-2015)  and regions $\mathcal{V}$  and $\mathcal{P}$. The mean of the extent for increments of average temperatures are shown in the last row.
				\label{TDecades2}}
			\begin{tabular}{l|cc|cc|cc|cc}
				\multicolumn{1}{l|}{$c$} & \multicolumn{4}{c}{$0^\circ$C} &  \multicolumn{4}{c}{$2^\circ$C}  	\\
				Decade &  \multicolumn{2}{c}{$D1$} &  \multicolumn{2}{c}{$D5$} & \multicolumn{2}{c}{$D1$} &  \multicolumn{2}{c}{$D5$}	\\
				Region & $\mathcal{V}$ & $\mathcal{P}$ & $\mathcal{V}$ & $\mathcal{P}$ & $\mathcal{V}$ & $\mathcal{P}$ & $\mathcal{V}$ & $\mathcal{P}$  \\
				\noalign{\smallskip}\hline\noalign{\smallskip}			
				$\{Y_{t,\ell}(\bs)-\tilde \mu(\bs)>0\}$ &  0.45 & 0.44 & 0.60 & 0.54 & 0.29 & 0.29 &  0.42 & 0.38 \\
				$\{Y_{t,\ell}(\bs)-\tilde \mu(\bs)>0; 2\}$ & 0.34 & 0.33 & 0.49 & 0.43 &  0.19 & 0.19 &  0.31  & 0.27 \\
				$\{Y_{t,\ell}(\bs)-\tilde \mu(\bs)>0; 3\}$	& 0.26 & 0.26 & 0.41 & 0.35 & 0.13 & 0.13 & 0.24 & 0.20 \\ 
				\noalign{\smallskip}\hline\noalign{\smallskip}
				$\{\bar Y_{D}(\bs)-\tilde \mu(\bs)>0\}$	&	  0.27 & 0.24 & 0.84 & 0.62  & 0.01 & 0.01 & 0.19 &0.10  \\
			\end{tabular}
		\end{center}
	\end{table}
	
	\begin{figure}[t]
		\begin{center}
			\includegraphics[width=0.45\textwidth]{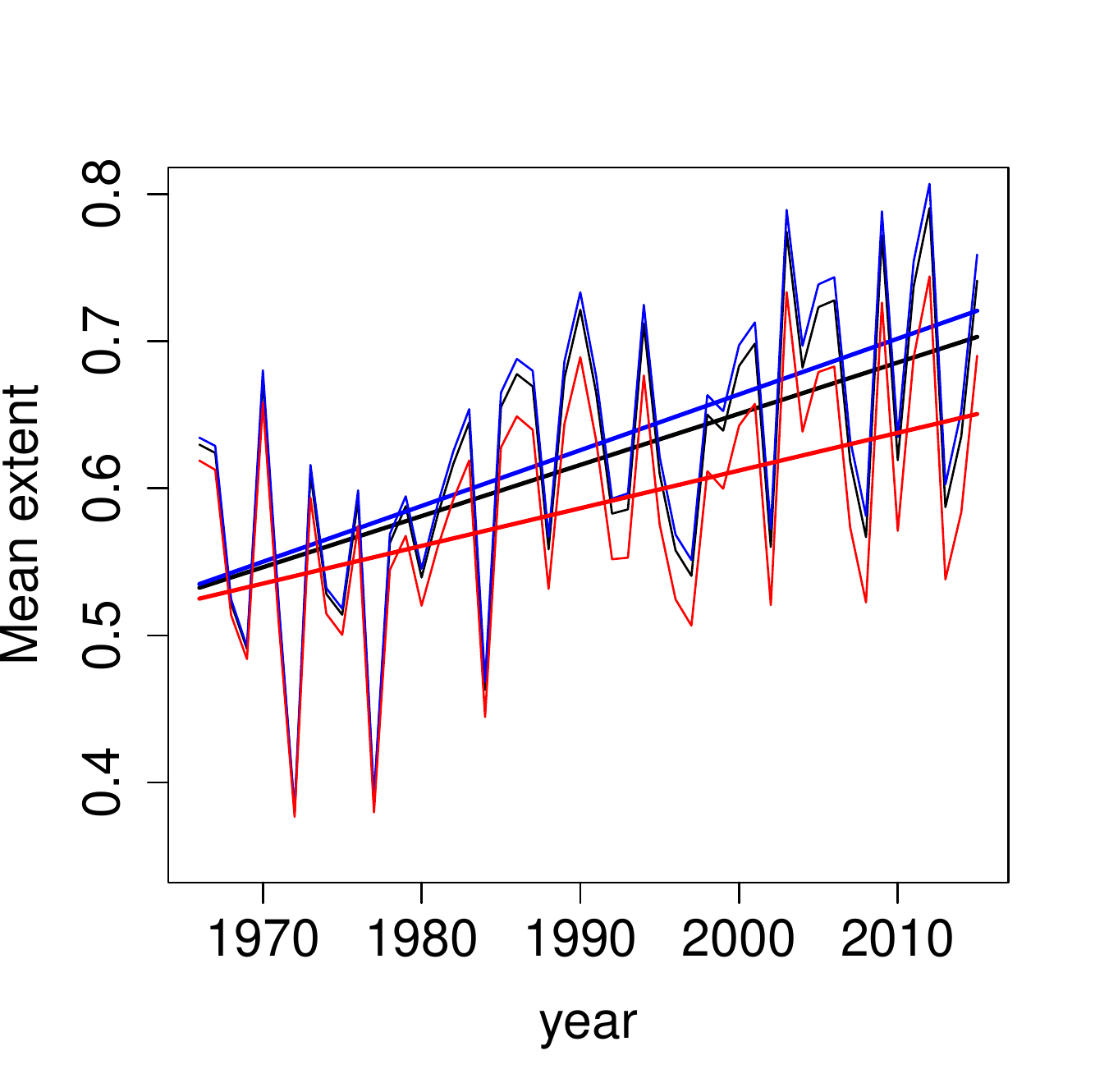}	
			\caption{Posterior mean across years  of the yearly average extents  for events  $\{ Y_{t,\ell}(\bs)-\tilde \mu(\bs)>0\}$  for $\mathcal{B}=\mathcal{D}$ (black), $\mathcal{V}$ (blue) and $\mathcal{P}$ (red), and corresponding regression lines.\label{F6reg}}
		\end{center}
	\end{figure}

	Figure \ref{F6reg} shows the  evolution over time  of the mean  of the  average extent in JJA in one year  for events $\{Y_{t,\ell}(\bs)-\tilde \mu(\bs)>0\}$ in $\mathcal{V}$,  $\mathcal{P}$, and in the entire region $\mathcal{D}$ for the sake of comparison. The corresponding  fitted  linear regressions are also plotted. A roughly linear increase   is observed in both regions, but with different trends, $0.0038$ and $0.0026$,   respectively; that means  an increase    in extent per decade of 3.8  in $\mathcal{V}$,  and 2.6 \%  in  $\mathcal{P}$.

\subsubsection{Extents for increments of average temperatures over  $\tilde \mu(\bs)$}

The last row in Table \ref{TDecades2} summarizes the mean  of the extents  for events $\{\bar Y_{D}(\bs)-\tilde \mu(\bs)>c\}$ with  $c=0,1,2^\circ$C.   In $D1$, the mean percentage of area with  average temperature higher than the reference mean is  quite similar in both regions, around 25\%. However, relevant differences appear in $D5$, where the mean percentage is 84 in $\mathcal{V}$ and 62\% in  $\mathcal{P}$.  
The posterior density of  the extents,  shown in  Fig.  S4 in the Supplement,  enables us to quantify the uncertainty of the extent and it  confirms the shift in location of the distribution  of the extent in $D5$ between the two regions. 
In $D1$, the posterior probability  of the extent for a positive  increment over $\tilde \mu(\bs)$ in  $\mathcal{V}$  being higher than  in $\mathcal{P}$  is 0.57 and  in $D5$, 0.94.

\subsubsection{Extents for  increments between average temperatures in $D1$ and $D5$}

Finally, we compare  the extent for  increments of average temperature, $\{\bar Y_{D5}(\bs)-\bar Y_{D1}(\bs)>c \}$ in  $\mathcal{V}$ and  $\mathcal{P}$. Table  \ref{TINcReg2} summarizes  the mean of those extents for $c=0,1,2^\circ$C. The  mean  percentage of area with a positive increment   is high in both regions, 98\% and 93\%,   respectively. However, there are differences in the extent of more strict events; e.g. the percentage of area with an increment higher than   $c=1^\circ$C   is 80\% in $\mathcal{V}$ and 63\% in $\mathcal{P}$.  The posterior density of  the extents,  shown in  Fig. S4 in the Supplement,  allows us to quantify the uncertainty. The posterior probability of the extent for a positive  increment in  $\mathcal{V}$  being higher than  in $\mathcal{P}$  is 0.96 and for increments higher than 1 and $2^\circ$C,  0.94 and 0.87, respectively.

To sum up, the increase in the extent of all the considered events is stronger in $\mathcal{V}$ than in $\mathcal{P}$, although the magnitude of the increase depends on the type of event.

\conclusions[Conclusions and future work]

\label{S5}

Acknowledging that climate change with regard to temperature is occurring both temporally and spatially,  we have presented tools to quantify this change.  
With regard to a collection of model-based replicates of spatially referenced time series of temperature data, the fundamental object is a daily event at an arbitrary location and time.  Working with exceedance events around the center of the temperature distribution, we offer two basic ideas: (i) probability surfaces which capture the spatial variation in the chance of an exceedance event and provide  climate risk  maps and (ii) extents which, for a subregion of interest, capture the expected proportion of incidence of a given exceedance event (as above) over the region.  Specifically, we define exceedance events in terms of a local mean value or  increments between two decades, but other definitions of interest could be used. These quantities are defined at daily scale and can be averaged to temporal scales of interest.  They are obtained from posterior predictive simulation of the collection of daily temperature series using a particular choice of hierarchical daily mean temperature model. 

Employing daily maximum temperature time series from $18$ spatial locations in Aragón (Spain) for more than $60$ years, comparison has been presented at daily and seasonal scale both temporally between decades and spatially between subregions. 
The analysis reveals  that there is an increase all over the region    in all the features and events used to quantify the evolution of temperature from 1966 to 2015. However,  that  increase is not spatially homogeneous, with   the largest increase arising  in the  center of the Ebro valley  and  NE area. The use of different events allows to quantify specific features, e.g., the probability of a daily temperature higher than the reference  mean has increased roughly 0.2 from  decade $D1$ (1966-1975) to $D5$ (2006-2015), attaining values higher than 0.7 in some areas in $D5$. As expected, the increase in features based on average temperatures is stronger: the probability of the average temperature being higher than the reference mean has increased  from $D1$ to $D5$ a value around 0.5,  being virtually 1 in some areas in $D5$.  In all of the region except a small NW area, the risk of  a positive  increment in the average temperatures between  $D1$ and $D5$ is virtually $1$, and the risk of an increase higher than $1^\circ$C is  higher than 0.5, and close to 1 in  the southcentral part of the valley and NE.

Concerning the spatial incidence in the entire region,  the extent  of daily temperature  higher than the reference  mean has  increased  3.5\% per decade. 
The mean of the extent with a positive average increment between $D1$ and $D5$ is 0.95. The suggested tools are also used to compare the evolution of the extent in  different regions. We found  that, in all the considered features, the  increase in  extent in region $\mathcal{V}$, with a semiarid climate, is higher than in $\mathcal{P}$, with a mountain climate. We showed that in $D1$  the extent of the average temperature  higher than the  reference mean is roughly 0.25 in  both regions while in $D5$ it increases to  0.84 in $\mathcal{V}$, and 0.62 in $\mathcal{P}$.  These results are  consistent with  those in  \cite{pena2021}  and \cite{haug2020spatial}, who  found a stronger change in summer mean temperatures in the center of the Ebro valley.

It is noteworthy that the suggested approach based on the analysis of  collection of spatially referenced time series of temperature data has important advantages versus the analysis of usually spatially scarce observed data. Although some empirical measures, such as the extent, could be  directly computed  from an observed dataset, in many cases,  they would be based on too few stations. Further, using only   observed series, even spatially dense gridded series, uncertainty cannot be  easily quantified, and probabilities or CIs such as those  suggested in this work, cannot be computed.

The proposed ideas can be applied to any spatio-temporal dataset collected over any region of interest.  All that is required is the fitting of a suitable model, retaining the model output for posterior prediction of temperatures.  Future work will investigate other regions of interest, making further comparisons.  We will also investigate events involving extremes of temperature, using suitable extreme value modeling, and compound events defined in terms of  maximum and minimum temperatures or other climate variables.  Further, we will explore other spatially referenced weather time series.  We also will attempt to forecast future temperature change, using a suitable version of our modeling, applied to climate scenarios.

\begin{acknowledgements}
This work was supported by the Grant PID2020-116873GB-I00 funded by MCIN/AEI/10.13039/501100011033; the Research Group E46\_20R: Modelos Estoc\'asticos funded by Gobierno de Arag\'on; and J. C.-M. was supported by the Doctoral Scholarship ORDEN CUS/581/2020 funded by Gobierno de Arag\'on.
The authors thank AEMET for providing the data. 
\end{acknowledgements}

\vspace*{-0.5cm}

\competinginterests{The authors declare that  no competing interests are present}

\vspace*{-0.5cm}

\codedataavailability{Code in R and temperature series used to fit the model are available upon request.} 


\bibliographystyle{copernicus} 
\bibliography{template.bib}    

\begin{thebibliography}{38}
\providecommand{\natexlab}[1]{#1}
\providecommand{\url}[1]{{\tt #1}}
\providecommand{\urlprefix}{URL }
\expandafter\ifx\csname urlstyle\endcsname\relax
  \providecommand{\doi}[1]{https://doi.org/\discretionary{}{}{}#1}\else
  \providecommand{\doi}{https://doi.org/\discretionary{}{}{}\begingroup
  \urlstyle{rm}\Url}\fi

\bibitem[{Angulo et~al.(1998)Angulo, Gonzalez-Manteiga, Febrero-Bande, and
  Alonso}]{Angulo98}
Angulo, J., Gonzalez-Manteiga, W., Febrero-Bande, M., and Alonso, F.:
  Semi-parametric statistical approaches for space-time process prediction,
  Environmental and Ecological Statistics, 5, 297--316, 1998.

\bibitem[{Banerjee et~al.(2014)Banerjee, Carlin, and Gelfand}]{BCG}
Banerjee, S., Carlin, B.~P., and Gelfand, A.~E.: Hierarchical Modeling and
  Analysis for Spatial Data, Chapman and Hall/CRC, New York, NY, USA, 2 edn.,
  \doi{10.1201/b17115}, 2014.

\bibitem[{Bolin and Lindgren(2015)}]{Bolin15}
Bolin, D. and Lindgren, F.: Excursion and contour uncertainty regions for
  latent {Gaussian} models, Journal of the Royal Statistical Society: Series B
  (Statistical Methodology), 77, 85--106, 2015.

\bibitem[{Caraway et~al.(2014)Caraway, McCreight, and Rajagopalan}]{Caraway14}
Caraway, N.~M., McCreight, J.~L., and Rajagopalan, B.: Multisite stochastic
  weather generation using cluster analysis and k-nearest neighbor time series
  resampling, J. Hydrol., 508, 197--213, 2014.

\bibitem[{Castillo-Mateo et~al.(2022)Castillo-Mateo, Lafuente, Asín, Cebrián,
  Gelfand, and Abaurrea}]{Castillo2022}
Castillo-Mateo, J., Lafuente, M., Asín, J., Cebrián, A.~C., Gelfand, A.~E.,
  and Abaurrea, J.: Spatial Modeling of Day-Within-Year Temperature Time
  Series: An Examination of Daily Maximum Temperatures in {A}rag\'on, {S}pain,
  Journal of Agricultural, Biological and Environmental Statistics, 27,
  487--505, \doi{10.1007/s13253-022-00493-3}, 2022.

\bibitem[{Cebri{\'a}n et~al.(2022)Cebri{\'a}n, As{\'\i}n, Gelfand, Schliep,
  Castillo-Mateo, Beamonte, and Abaurrea}]{cebrian2021spatio}
Cebri{\'a}n, A.~C., As{\'\i}n, J., Gelfand, A.~E., Schliep, E.~M.,
  Castillo-Mateo, J., Beamonte, M.~A., and Abaurrea, J.: Spatio-temporal
  analysis of the extent of an extreme heat event, Stochastic Environmental
  Research and Risk Assessment, 36, 2737--2751,
  \doi{10.1007/s00477-021-02157-z}, 2022.

\bibitem[{Craigmile and Guttorp(2011)}]{Craigmile11}
Craigmile, P.~F. and Guttorp, P.: Space-time modelling of trends in temperature
  series, Journal of Time Series Analysis, 32, 378--395,
  \doi{10.1111/j.1467-9892.2011.00733.x}, 2011.

\bibitem[{Dowlatabadi and Morgan(1993)}]{Dowlatabadi93}
Dowlatabadi, H. and Morgan, M.~G.: Integrated assessment of climate change,
  Science, 259, 1813--1932, 1993.

\bibitem[{French(2017)}]{French17}
French, J.~P.: autoimage: Multiple Heat Maps for Projected Coordinates, The R
  Journal, 9, 284--297, 2017.

\bibitem[{Garc{\'\i}a-Valero et~al.(2015)Garc{\'\i}a-Valero, Mont{\'a}vez,
  G{\'o}mez-Navarro, and Jim{\'e}nez-Guerrero}]{Garcia15}
Garc{\'\i}a-Valero, J.~A., Mont{\'a}vez, J.~P., G{\'o}mez-Navarro, J., and
  Jim{\'e}nez-Guerrero, P.: Attributing trends in extremely hot days to changes
  in atmospheric dynamics, Natural Hazards and Earth System Sciences, 15,
  2143--2159, 2015.

\bibitem[{Gelfand and Smith(1990)}]{Gelfand90}
Gelfand, A.~E. and Smith, A.~F.: Sampling-based approaches to calculating
  marginal densities, Journal of the American statistical association, 85,
  398--409, 1990.

\bibitem[{Hartfield and Gunst(2003)}]{Hartfield03}
Hartfield, M.~I. and Gunst, R.~F.: Identification of model components for a
  class of continuous spatiotemporal models, Journal of agricultural,
  biological, and environmental statistics, 8, 105--121, 2003.

\bibitem[{Hartmann et~al.(2013)Hartmann, Klein~Tank, Rusticucci, Alexander,
  Br{\"o}nnimann, Charabi, Dentener, Dlugokencky, Easterling, Kaplan
  et~al.}]{hartmann2013climate}
Hartmann, D., Klein~Tank, A., Rusticucci, M., Alexander, L., Br{\"o}nnimann,
  S., Charabi, Y., Dentener, F., Dlugokencky, E., Easterling, D., Kaplan, A.,
  et~al.: {Climate Change 2013: The Physical Science Basis. Contribution of
  Working Group I to the Fifth Assessment Report of the Intergovernmental Panel
  on Climate Change}, Observations: Atmosphere and Surface, edited by T.
  Stocker, D. Qin, G.-K. Plattner, M. Tignor, S. Allen, J. Boschung, A. Nauels,
  Y. Xia, V. Bex, and P. Midgley (Cambridge University Press, 2013), 2013.

\bibitem[{Haug et~al.(2020)Haug, Thorarinsdottir, S{\o}rbye, and
  Franzke}]{haug2020spatial}
Haug, O., Thorarinsdottir, T.~L., S{\o}rbye, S.~H., and Franzke, C.~L.: Spatial
  trend analysis of gridded temperature data at varying spatial scales,
  Advances in Statistical Climatology, Meteorology and Oceanography, 6, 1--12,
  2020.

\bibitem[{IPCC(2018)}]{IPCC2018}
IPCC: {Summary for Policymakers}, in: {Global warming of 1.5$^{\circ}${C}. {A}n
  {IPCC} {S}pecial {R}eport on the impacts of global warming of
  1.5$^{\circ}${C} above pre-industrial levels and related global greenhouse
  gas emission pathways, in the context of strengthening the global response to
  the threat of climate change, sustainable development, and efforts to
  eradicate poverty}, edited by Masson-Delmotte, V., Zhai, P., P{\"o}rtner,
  H.-O., Roberts, D., Skea, J., Shukla, P.~R., Pirani, A., Moufouma-Okia, W.,
  P{\'e}an, C., Pidcock, R., Connors, S., Matthews, J. B.~R., Chen, Y., Zhou,
  X., Gomis, M.~I., Lonnoy, E., Maycock, T., Tignor, M., and Waterfield, T.,
  World Meteorological Organization, Geneva, Switzerland, 2018.

\bibitem[{Katz(2002)}]{Katz02}
Katz, R.~W.: Techniques for estimating uncertainty in climate change scenarios
  and impact studies, Climate research, 20, 167--185, 2002.

\bibitem[{Keellings and Moradkhani(2020)}]{keellings2020spatiotemporal}
Keellings, D. and Moradkhani, H.: {Spatiotemporal evolution of heat wave
  severity and coverage across the United States}, Geophysical Research
  Letters, 47, e2020GL087\,097, 2020.

\bibitem[{Khan et~al.(2019)Khan, Shahid, Ismail, Ahmed, and
  Nawaz}]{khan2019trends}
Khan, N., Shahid, S., Ismail, T., Ahmed, K., and Nawaz, N.: {Trends in heat
  wave related indices in Pakistan}, Stochastic environmental research and risk
  assessment, 33, 287--302, 2019.

\bibitem[{Kleiber et~al.(2013)Kleiber, Katz, and Rajagopalan}]{Kleiber13}
Kleiber, W., Katz, R.~W., and Rajagopalan, B.: Daily minimum and maximum
  temperature simulation over complex terrain, The Annals of Applied
  Statistics, 7, 588--612, 2013.

\bibitem[{Li and Thompson(2021)}]{Li21}
Li, J. and Thompson, D.~W.: Widespread changes in surface temperature
  persistence under climate change, Nature, 599, 425--430, 2021.

\bibitem[{Lyon et~al.(2019)Lyon, Barnston, Coffel, and
  Horton}]{lyon2019projected}
Lyon, B., Barnston, A.~G., Coffel, E., and Horton, R.~M.: {Projected increase
  in the spatial extent of contiguous US summer heat waves and associated
  attributes}, Environmental Research Letters, 14, 114\,029, 2019.

\bibitem[{Masson-Delmotte et~al.(2021)Masson-Delmotte, Zhai, Pirani, Connors,
  P{\'e}an, Berger, Caud, Chen, Goldfarb, Gomis et~al.}]{masson2021climate}
Masson-Delmotte, V., Zhai, P., Pirani, A., Connors, S.~L., P{\'e}an, C.,
  Berger, S., Caud, N., Chen, Y., Goldfarb, L., Gomis, M.~I., et~al.: {Climate
  Change 2021: The Physical Science Basis. Contribution of Working Group I to
  the Sixth Assessment Report of the Intergovernmental Panel on Climate
  Change}, IPCC: Geneva, Switzerland, 2021.

\bibitem[{Pe{\~n}a-Angulo et~al.(2021)Pe{\~n}a-Angulo, Gonzalez-Hidalgo,
  Sandon{\'i}s, Beguer{\'i}a, Tomas-Burguera, L{\'o}pez-Bustins, Lemus-Canovas,
  and Martin-Vide}]{pena2021}
Pe{\~n}a-Angulo, D., Gonzalez-Hidalgo, J.~C., Sandon{\'i}s, L., Beguer{\'i}a,
  S., Tomas-Burguera, M., L{\'o}pez-Bustins, J.~A., Lemus-Canovas, M., and
  Martin-Vide, J.: Seasonal temperature trends on the {S}panish mainland: {A}
  secular study (1916--2015), International Journal of Climatology, 41,
  3071--3084, \doi{10.1002/joc.7006}, 2021.

\bibitem[{Pfleiderer and Coumou(2018)}]{pfleiderer2018quantification}
Pfleiderer, P. and Coumou, D.: {Quantification of temperature persistence over
  the Northern Hemisphere land-area}, Climate Dynamics, 51, 627--637, 2018.

\bibitem[{Rebetez et~al.(2009)Rebetez, Dupont, and
  Giroud}]{rebetez2009analysis}
Rebetez, M., Dupont, O., and Giroud, M.: {An analysis of the July 2006 heatwave
  extent in Europe compared to the record year of 2003}, Theoretical and
  Applied Climatology, 95, 1--7, 2009.

\bibitem[{Schliep et~al.(2021)Schliep, Gelfand, Abaurrea, As{\'\i}n, Beamonte,
  and Cebri{\'a}n}]{schliep2021long}
Schliep, E.~M., Gelfand, A.~E., Abaurrea, J., As{\'\i}n, J., Beamonte, M.~A.,
  and Cebri{\'a}n, A.~C.: Long-term spatial modelling for characteristics of
  extreme heat events, Journal of the Royal Statistical Society: Series A
  (Statistics in Society), 184, 1070--1092, 2021.

\bibitem[{Scott and Chandler(2011)}]{scott2011statistical}
Scott, M. and Chandler, R.: Statistical methods for trend detection and
  analysis in the environmental sciences, John Wiley \& Sons, 2011.

\bibitem[{Smith et~al.(2018)Smith, Strong, and Rassoul-Agha}]{Smith18}
Smith, K., Strong, C., and Rassoul-Agha, F.: Multisite generalization of the
  SHArP weather generator, Journal of Applied Meteorology and Climatology, 57,
  2113--2127, 2018.

\bibitem[{Sommerfeld et~al.(2018)Sommerfeld, Sain, and
  Schwartzman}]{Sommerfeld18}
Sommerfeld, M., Sain, S., and Schwartzman, A.: Confidence regions for spatial
  excursion sets from repeated random field observations, with an application
  to climate, Journal of the American Statistical Association, 113, 1327--1340,
  2018.

\bibitem[{Stroud et~al.(2001)Stroud, Muller, and Sanso}]{Stroud01}
Stroud, J.~R., Muller, P., and Sanso, B.: Dynamic Models for Spatiotemporal
  Data, Journal of the Royal Statistical Society. Series B (Statistical
  Methodology), 63, 673--689, 2001.

\bibitem[{Thorarinsdottir et~al.(2017)Thorarinsdottir, Guttorp, Drews,
  Kaspersen, and de~Bruin}]{Thorarinsdottir2017}
Thorarinsdottir, T., Guttorp, P., Drews, M., Kaspersen, P.~S., and de~Bruin,
  K.: Sea level adaptation decisions under uncertainty, Water Resources
  Research, 53, 8147--8163, 2017.

\bibitem[{Tye et~al.(2019)Tye, Katz, and Rajagopalan}]{tye2019climate}
Tye, M.~R., Katz, R.~W., and Rajagopalan, B.: {Climate change or climate
  regimes? Examining multi-annual variations in the frequency of precipitation
  extremes over the Argentine Pampas}, Climate dynamics, 53, 245--260, 2019.

\bibitem[{Verdin et~al.(2019)Verdin, Rajagopalan, Kleiber, Podest{\'a}, and
  Bert}]{verdin2019baygen}
Verdin, A., Rajagopalan, B., Kleiber, W., Podest{\'a}, G., and Bert, F.:
  {BayGEN}: {A} {B}ayesian space-time stochastic weather generator, Water
  Resources Research, 55, 2900--2915, 2019.

\bibitem[{Wikle et~al.(2001)Wikle, Milliff, Nychka, and Berliner}]{Wikle01}
Wikle, C.~K., Milliff, R.~F., Nychka, D., and Berliner, L.~M.: {Spatiotemporal
  hierarchical Bayesian modeling tropical ocean surface winds}, Journal of the
  American Statistical Association, 96, 382--397, 2001.

\bibitem[{Wilks(1999)}]{Wilks99}
Wilks, D.~S.: Simultaneous stochastic simulation of daily precipitation,
  temperature and solar radiation at multiple sites in complex terrain, Agric.
  For. Meteor., 96, 85--101, 1999.

\bibitem[{Wilks(2009)}]{Wilks09}
Wilks, D.~S.: A gridded multisite weather generator and synchronization to
  observed weather data, Water Resour. Res., 45, W10\,419, 2009.

\bibitem[{Zscheischler et~al.(2020)Zscheischler, Martius, Westra, Bevacqua,
  Raymond, Horton, van~den Hurk, AghaKouchak, J{\'e}z{\'e}quel, Mahecha
  et~al.}]{zscheischler2020typology}
Zscheischler, J., Martius, O., Westra, S., Bevacqua, E., Raymond, C., Horton,
  R.~M., van~den Hurk, B., AghaKouchak, A., J{\'e}z{\'e}quel, A., Mahecha,
  M.~D., et~al.: A typology of compound weather and climate events, Nature
  reviews earth \& environment, 1, 333--347, 2020.

\bibitem[{Zwiers and Von~Storch(1995)}]{zwiers1995taking}
Zwiers, F.~W. and Von~Storch, H.: Taking serial correlation into account in
  tests of the mean, Journal of Climate, 8, 336--351, 1995.

\end{thebibliography}


\begin{thebibliography}{2}
\providecommand{\natexlab}[1]{#1}
\providecommand{\url}[1]{{\tt #1}}
\providecommand{\urlprefix}{URL }
\expandafter\ifx\csname urlstyle\endcsname\relax
  \providecommand{\doi}[1]{https://doi.org/\discretionary{}{}{}#1}\else
  \providecommand{\doi}{https://doi.org/\discretionary{}{}{}\begingroup
  \urlstyle{rm}\Url}\fi

\bibitem[{Castillo-Mateo et~al.(2022)Castillo-Mateo, Lafuente, Asín, Cebrián,
  Gelfand, and Abaurrea}]{Castillo2022}
Castillo-Mateo, J., Lafuente, M., Asín, J., Cebrián, A.~C., Gelfand, A.~E.,
  and Abaurrea, J.: Spatial Modeling of Day-Within-Year Temperature Time
  Series: An Examination of Daily Maximum Temperatures in {A}rag\'on, {S}pain,
  Journal of Agricultural, Biological and Environmental Statistics, 27,
  487--505, \doi{10.1007/s13253-022-00493-3}, 2022.

\bibitem[{Zhang(2004)}]{zhang2004}
Zhang, H.: Inconsistent estimation and asymptotically equal interpolations in
  model-based geostatistics, Journal of the American Statistical Association,
  99, 250--261, \urlprefix\url{https://doi.org/10.1198/016214504000000241},
  2004.

\end{thebibliography}

\end{document}


\title{ Supplement of the paper: Model-based tools for assessing space and time change in daily maximum temperature: an application to the Ebro basin in Spain
\vspace*{1cm}}

\Author[1]{Ana C.}{ Cebri\'an}

\Author[1]{Jes\'us}{ As\'in}

\Author[1]{Jorge}{Castillo-Mateo}

\Author[2]{Alan E.}{Gelfand}

\Author[1]{Jes\'us}{Abaurrea }

\affil[1]{Department of Statistical Methods, University of Zaragoza, Pedro Cerbuna 12, Zaragoza 50009, Spain}

\affil[2]{Department of Statistical Science, Duke University, Durham NC 27710, USA}

\runningtitle{Supplement: Model-based tools for assessing space and time change in daily temperature}

\runningauthor{TEXT}

\maketitle

This supplement includes  some complementary information on the  point-referenced hierarchical model  presented in \cite{Castillo2022}, that it is used to generate  the collection of daily temperature series at geo-coded locations in the central Ebro  basin (Spain). Those generated series and the tools suggested in the work  are employed to analyze temperature evolution in  that region.

\section{Description of  the data set} \label{SupExplor}

\begin{figure}
\begin{center}
	\includegraphics[width=0.95\linewidth]{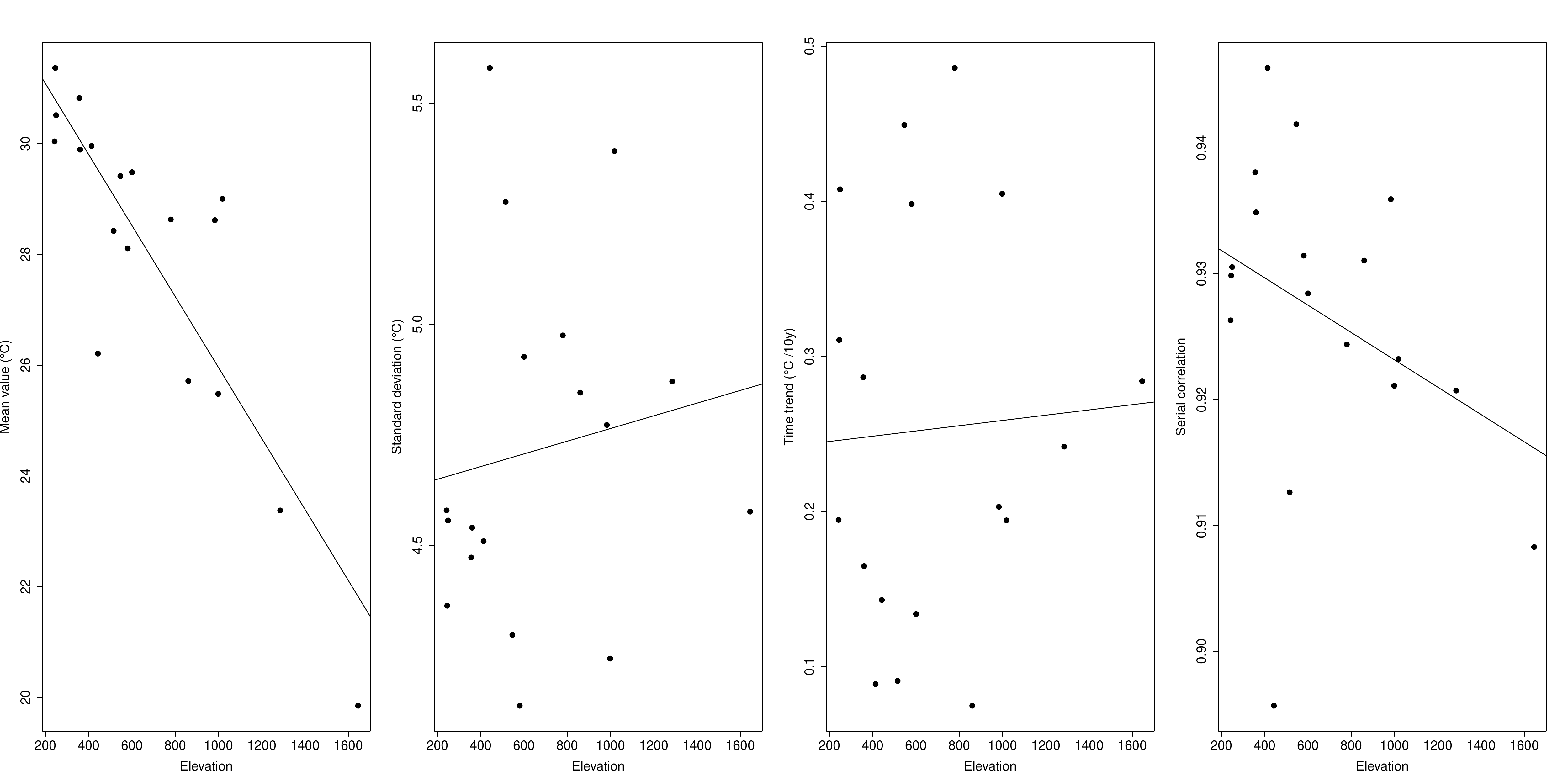}
\caption{Mean, standard deviation, linear trend ($^{\circ}\text{C}/\text{decade}$) and serial correlation  of daily maximum temperature in JJA 1956-2015 versus elevation of 18 locations in the  Ebro basin. 
 \label{FigObsDescripElevAppendix} }
\end{center}
\end{figure}

Table 	\ref{TableObsSummaryAppendix}  summarizes the mean, the standard deviation,  the linear trend and  the serial correlation of the daily maximum temperature series in JJA  for 1956-2015 at  18 locations  in the  Ebro basin (Spain). Figure  \ref{FigObsDescripElevAppendix} shows those four summary  measures   versus elevation.

\begin{table}[tb]
	\begin{center}
		\caption{Mean, standard deviation,  linear trend ($^{\circ}\text{C}/\text{decade}$) and   serial correlation  of daily maximum temperature in JJA 1956-2015
			\label{TableObsSummaryAppendix}}
		\begin{tabular}{|l|c|c|c|c|} \hline
Station&Mean value  ($^{\circ}\text{C}$)&Std dev&Linear trend  & Serial correlation \\ \hline
Pamplona       &26.2&5.6&0.1&0.90\\ \hline
Ansó           &25.7&4.8&0.1&0.93\\ \hline
Sallent        &23.4&4.9&0.2&0.92\\ \hline
Panticosa      &19.9&4.6&0.3&0.91\\ \hline
El Bayo        &29.9&4.5&0.2&0.93\\ \hline
Yesa           &28.4&5.3&0.1&0.91\\ \hline
La Sotonera    &30.0&4.5&0.1&0.95\\ \hline
Huesca         &29.4&4.3&0.4&0.94\\ \hline
Bu\~nuel         &30.0&4.6&0.2&0.93\\ \hline
Zaragoza       &30.5&4.6&0.4&0.93\\ \hline
Pallaruelo     &30.8&4.5&0.3&0.94\\ \hline
Puebla de Híjar&31.4&4.4&0.3&0.93\\ \hline
Calatayud      &29.5&4.9&0.1&0.93\\ \hline
Tornos         &29.0&5.4&0.2&0.92\\ \hline
Daroca         &28.6&5.0&0.5&0.92\\ \hline
Cueva Foradada &28.1&4.1&0.4&0.93\\ \hline
Sta. Eulalia   &28.6&4.8&0.2&0.94\\ \hline
Morella        &25.5&4.2&0.4&0.92\\ \hline
		\end{tabular}
	\end{center}
\end{table}

\vfill 
\eject

\section{Prior specification of the space-time model}
\label{S1}

With respect to the prior distribution of the parameters in the model by \cite{Castillo2022},  diffuse and, when available, conjugate prior distributions are selected.  The intercept and all four slope fixed coefficients are independent normal with zero mean and standard deviation 100.  All seven variance parameters are independent $Inverse$-$Gamma(0.1,0.1)$.  For identificability, the random effect for the first year, $\psi_1$, is fixed to zero.  Also, $Z_{\rho}$ and $Z_{\sigma}$ are normal with zero mean and standard deviation 100 and 1, respectively.  Finally, all four decay parameters are fixed to $3 / d_{max}$, where $d_{max}$ is the maximum distance between any pair of spatial locations. That is, with an exponential covariance function, the decay parameter is $3/\text{range}$, and it is set to the value associated with the largest spatial range of the data observed.  This is motivated by two aspects, the preliminary analysis developed in \cite{Castillo2022} and the fact that, with an exponential covariance function, the variance and the decay parameter are not individually identified \citep{zhang2004}.

\section{Fitted space-time model}

\begin{figure}[t]
	\begin{center}
		\includegraphics[width=0.45\textwidth]{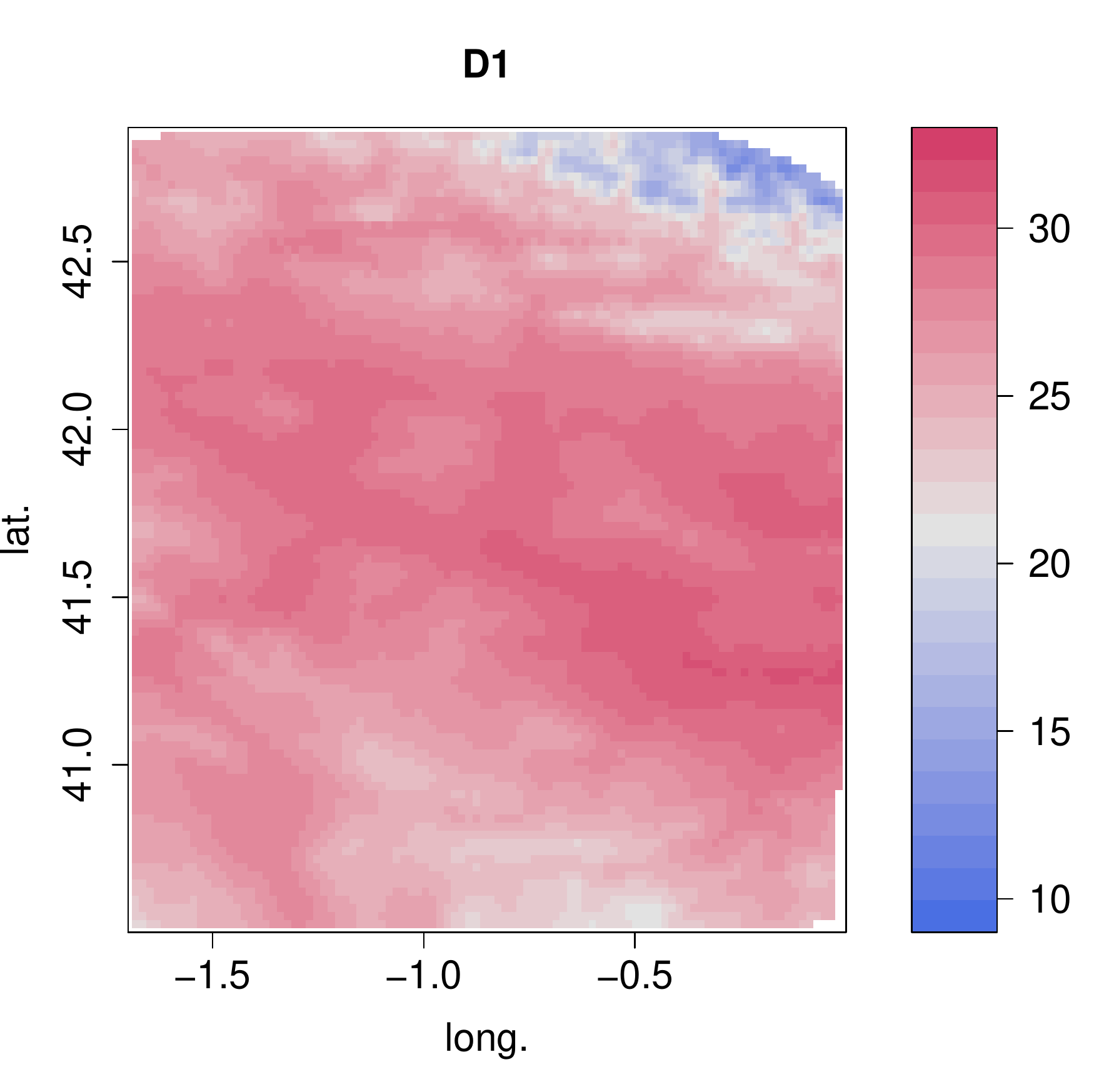}	
		\includegraphics[width=0.45\textwidth]{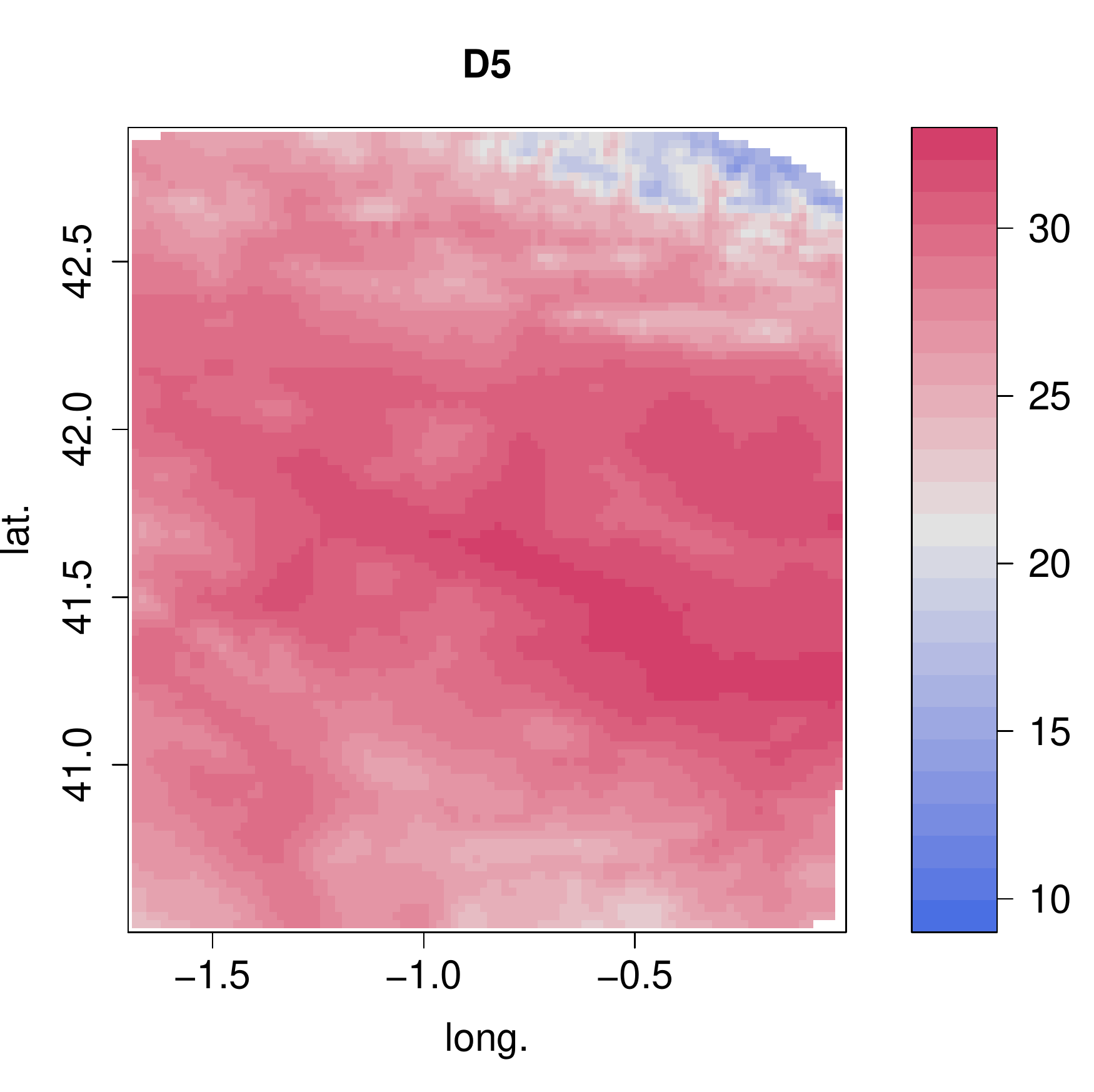}
		\end{center}
	\caption{Median of the daily temperatures  simulated  by the model in JJA in decades $D1$ (left) and $D5$  (right). \label{FP} }
\end{figure}

As a simple example  of the information provided by the output series from the fitted model, Figure \ref{FP} shows, spatially, the medians  of temperature series in JJA  in decades  $D1$  and $D5$.

\section{More results for the space-time analysis}

\subsection{Analysis  of daily increments between $D1$ and $D5$}

Here, we seek to define events to evaluate  the increment  between those decades at a daily scale, capturing the  variability of daily temperatures. To that end, we characterize the difference between the temperature for one day in $D1$ and the corresponding day in $D5$. We define $Z_{j,\ell}(\bs)=Y_{2005+j,\ell}(\bs)-Y_{1965+j,\ell}(\bs)$,  the difference  between temperature  at the same day $\ell$  in year $1965+j$, in $D1$,  and year $2005+j$, in $D5$,   for $\ell$ in JJA, and  $j=1, \ldots, 10$.  
The advantage  of using the differences  $Z_{j,\ell}(\bs)$ is that  the  seasonal effect  of daily temperatures is canceled so  we capture changes across years.

We consider events where  the daily  increment is larger than  a constant $c$, $ \{Z_{j,\ell}(\bs) >c\}$   with $c=0, 1$ and $2^\circ$C. To analyze the persistence of the increments across days, we define events  based  on the differences of $k=2$ or 3  consecutive days $\{Z_{j,\ell}(\bs) >c; 2\} \equiv \{Z_{j,\ell}(\bs), Z_{j,\ell+1}(\bs) >c\}$ or $\{Z_{j,\ell}(\bs) >c; 3\} \equiv \{ Z_{j,\ell-1}(\bs), Z_{j,\ell}(\bs), Z_{j,\ell+1}(\bs) >c\}$. 

Note that  the difference of average temperatures $\bar Y_{D5}(\bs)-\bar Y_{D1}(\bs)$ is equal to the average of the daily increments  in  JJA in one decade,
$$\bar Z(\bs)=\frac{1}{ 920}\sum_ {j=1}^{10}  \sum_ {\ell \in JJA} Z_{j,\ell}(\bs).$$
Then, the events  $\{\bar Y_{D5}(\bs)-\bar Y_{D1}(\bs)>c\}$  in the main text  can also be defined as  $\{\bar Z(\bs) >c\}$.

\paragraph{Surface of  probabilities}

\begin{figure}[t]
	\begin{center}
		\includegraphics[width=0.32\textwidth]{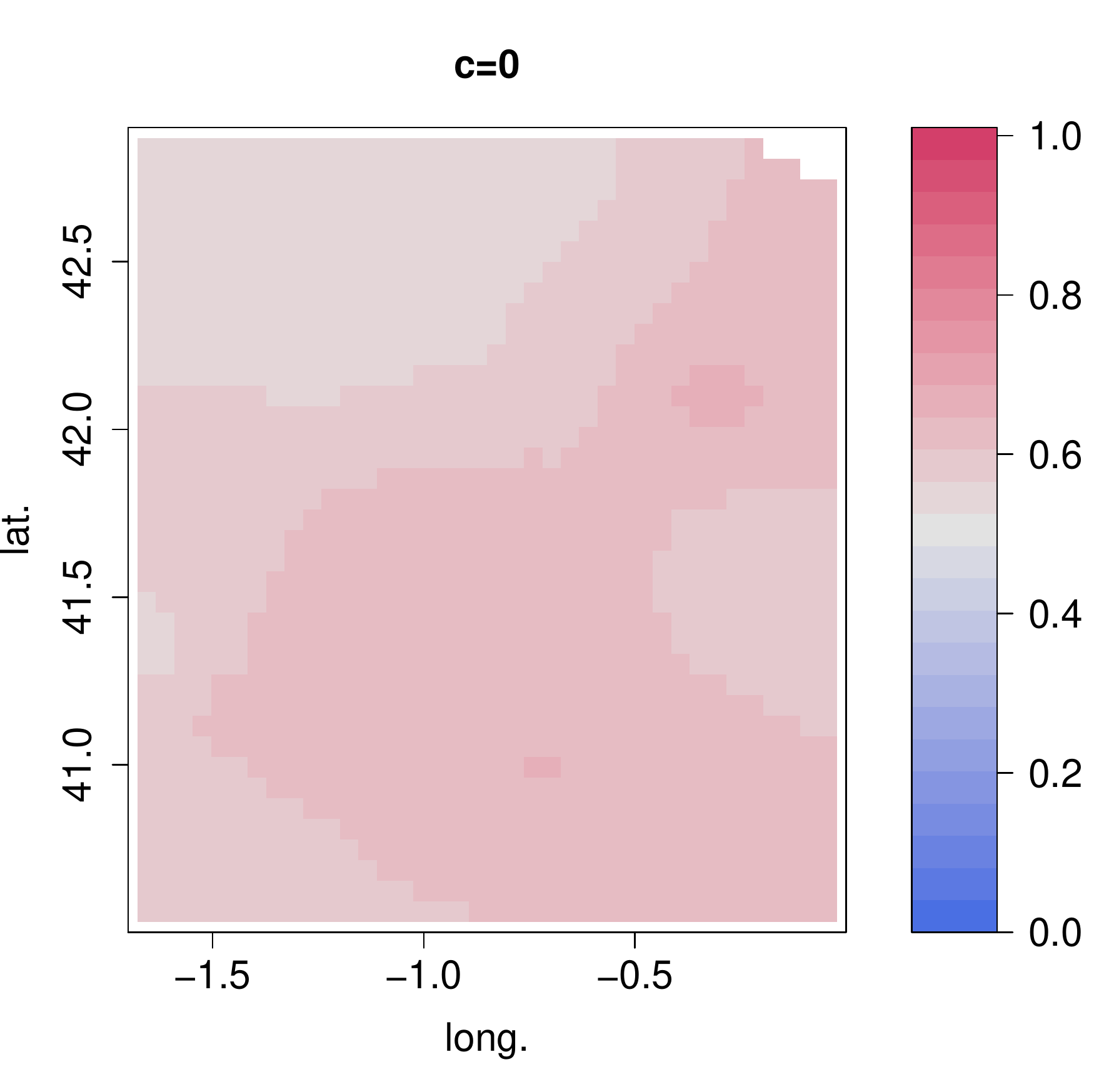}
		\includegraphics[width=0.32\textwidth]{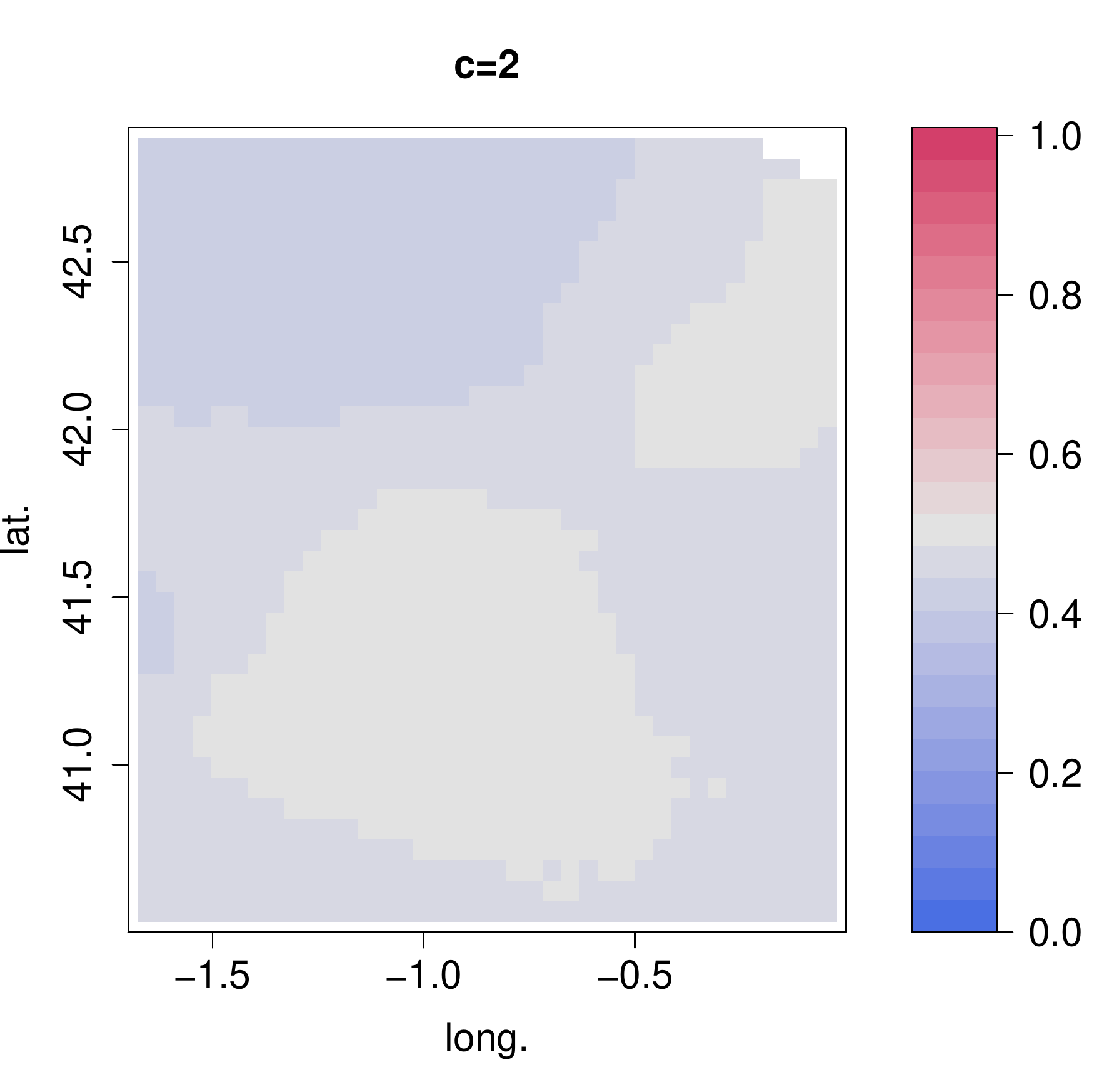}
		\includegraphics[width=0.32\textwidth]{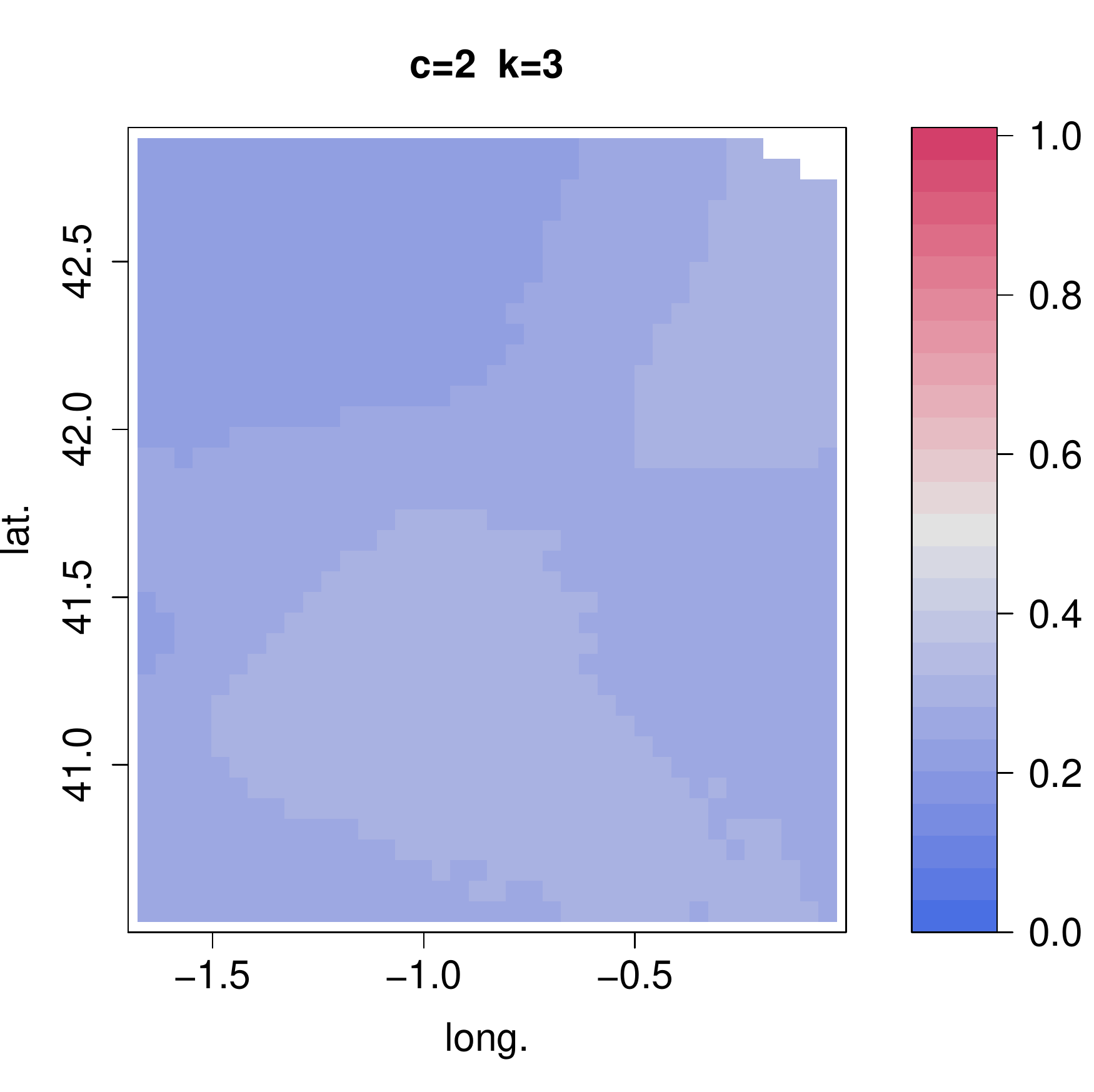}		
	\end{center}
	\caption{Average over JJA and index $j$ of the posterior probabilities of  events  $\{Z_{j,\ell}(\bs)>0\}$ (left),  $\{Z_{j,\ell}(\bs)>2\}$ (middle) and persistent events $\{Z_{j,\ell}(\bs)>2; 3\}$ (right). \label{F4b}}
\end{figure}

We compute   the  posterior probabilities for the events  $ \{Z_{j,\ell}(\bs) >c\}$, with $c=0,1,2^\circ$C,   and  we average them  over days and index $j$,
$ \frac{1}{920}  	\sum_{j=1}^{10}	\sum_{\ell \in JJA} \tilde P\left(Z_{j,\ell}(\bs) >c \right)$.
 We also compute the probabilities for the persistent events  defined with  $k=2$ and $3$ days; Figure   \ref{F4b}  shows  the average probabilities  for $\{Z_{j,\ell}(\bs)>0\}$,  $\{Z_{j,\ell}(\bs)>2\}$ and  $\{Z_{j,\ell}(\bs)>2; 3\}$. Remarkably, the spatial variability of these probabilities is quite  low and no relevant differences are observed: the highest  values are observed in the central south part of the basin and in the northeast, but the differences with other areas are smaller than 0.1 for the three events.

\paragraph{Average extents based on increments of daily temperatures}

The  extent for the  events  based on increments of daily temperature  are summarized by calculating  averages over days and index $j$,
$\frac{1}{920}\sum_{j=1}^{10} \  \sum_{\ell\in JJA} \widetilde{Ext}\left( Z_{j,\ell}(\bs) > c \right)$.

Table  \ref{TMCI}  summarizes the posterior mean of these average extents  for  events $\{Z_{j,\ell}(\bs)>c\}$ and $\{Z_{j,\ell}(\bs)>c;k\}$  for  $c=0,1,2$ and $k=2,3$. The mean   and the 90\% CI of the average extent  with a positive  increase  between $D1$ and $D5$  is  $0.60 \ (0.59,0.61)$,  and  with an increment higher than $2^\circ$C, $0.47\ (0.46, 0.48)$. 

The  corresponding values for  persistent events are lower but still quite high: in mean, the average percentage of area with  a  positive increment  in  three days in a row is 39\%,  and with an increment larger than $2^\circ$C,  27\%.

\begin{table}[t]
	\begin{center}
		\caption{Posterior mean of the average extent   for events   based on daily increments in the entire region and in regions $\mathcal{B}1$ and  $\mathcal{B}2$.}
		\label{TMCI}
		\begin{tabular}{l|ccc|ccc|ccc}
			Event & \multicolumn{3}{c|}{$\{Z_{j,\ell}(\bs)>c\}$ } & \multicolumn{3}{c|}{$\{Z_{j,\ell}(\bs)>c; 2\}$ }  & \multicolumn{3}{c}{$\{Z_{j,\ell}(\bs)>c; 3\}$ }  \\  
			$c$	& 0 & 1 & 2 & 0 & 1 & 2 & 0 & 1 & 2 \\ \hline
			$\mathcal{D}$ &  0.60 &  0.53 &  0.47 & 0.48& 0.41& 0.35 & 0.39 &0.33 & 0.27 \\	
			$\mathcal{B}1$ &  0.61 & 0.54 & 0.48 & 0.49 & 0.42 & 0.35 & 0.40 & 0.33 & 0.27 \\	
            $\mathcal{B}2$& 0.57 & 0.51 & 0.45 & 0.45 & 0.39 & 0.33 & 0.37 & 0.31 & 0.25 \\
		\end{tabular}
	\end{center}
\end{table}

\paragraph{Average extent  of daily increments in areas with different climates}

To compare the extent  for daily  events, Table \ref{TMCI}  summarizes the posterior mean of the  average extent for events based on  daily increments between $D5$ and $D1$ $ \{ Z_{j,\ell}(\bs)>c; k \} $  for  $c=0,1,2$ and $k=1,2,3$ in  both regions. The mean of the  average  percentage of area with a positive  increase  is  slightly higher in region  $\mathcal{B}1$  than in  $\mathcal{B}2$ with values   61 and 57\%,  respectively.  A similar difference between regions is observed in  all the types of events; e.g., the mean of the average percentage of area with a positive increase in  three days in a row is 40 and 37\%, respectively. Even the  mean of the percentage of area with increments higher than $2^\circ$C in three days is not negligible, with values higher than 25\% in both regions.

\subsection{Posterior density of the extent  for some events in areas with different climates}

\begin{figure}[H]
	\begin{center}
		\includegraphics[width=0.4\textwidth]{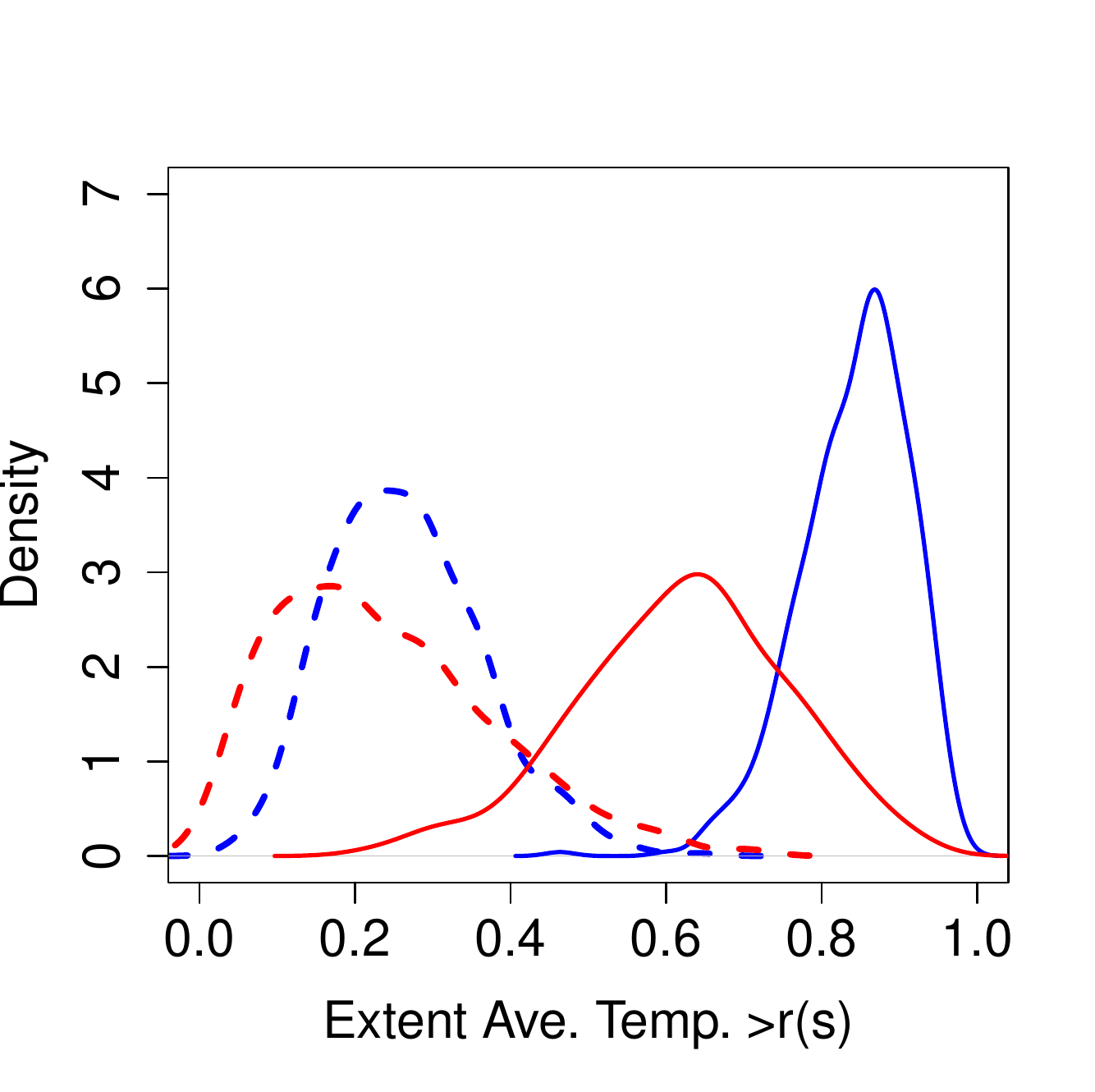}
		\includegraphics[width=0.4\textwidth]{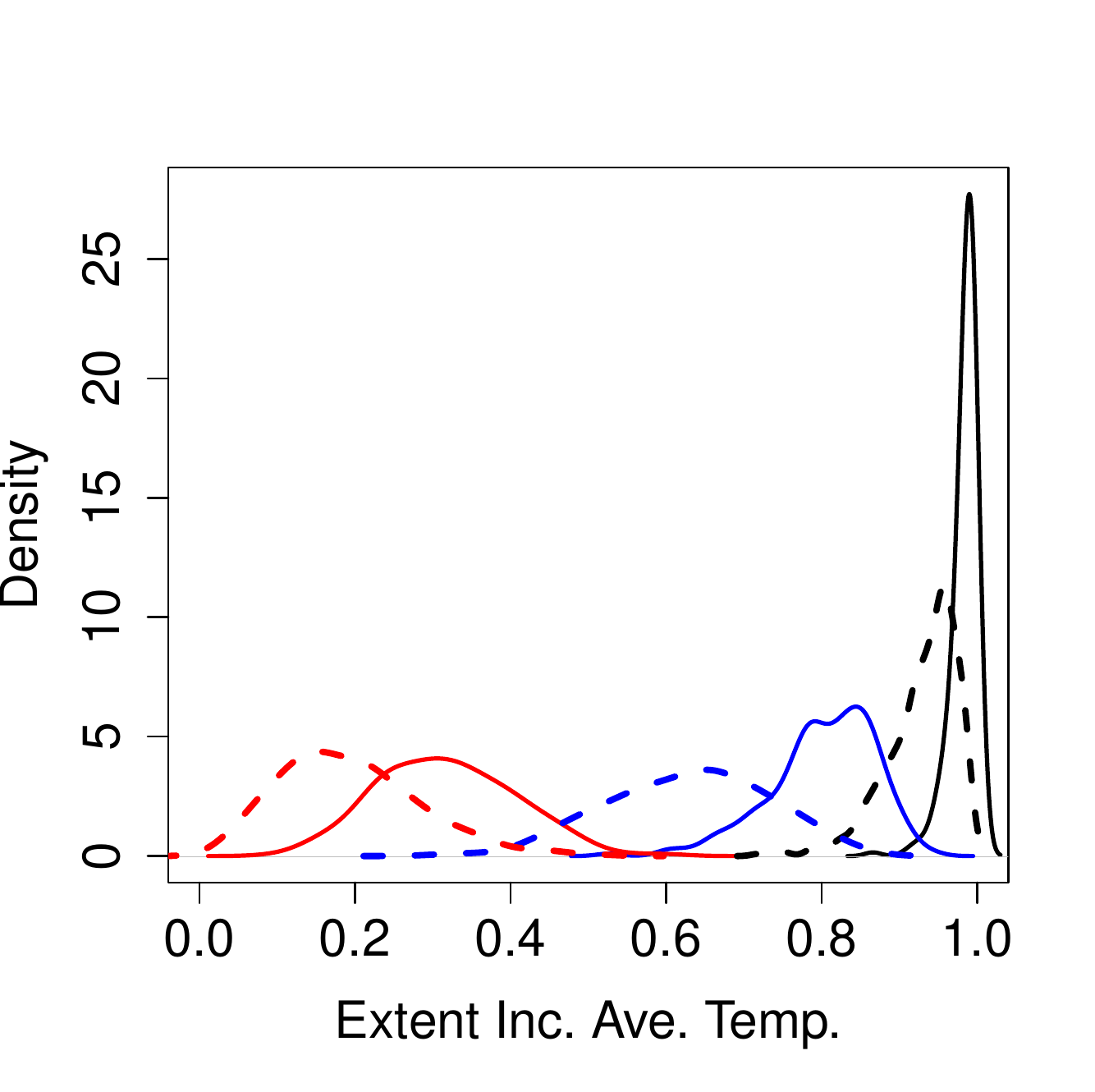}	
	\end{center}		
	\caption{Left: Posterior density of  $\widetilde{Ext}\left( \bar Y_{D}(\bs) - \mu(\bs)>0; \mathcal{B}\right)$, for $\mathcal{B}1$  (solid line) and $\mathcal{B}2$  (dotted line) and  $D1$ (red) $D5$ (blue). Right: Posterior density of     $\widetilde{Ext}\left(\bar Y_{D5}(\bs)-\bar Y_{D1}(\bs) > c;\mathcal{B} \right)$ for $c=0,1,2^\circ$C  (black, blue and red) in $\mathcal{B}1$  (solid line) and $\mathcal{B}2$  (dotted line). 
	\label{F6c} }
\end{figure}

 Figure \ref{F6c} (left) shows the posterior density of $\widetilde{Ext} \left(\bar Y_{D}(\bs)-\tilde \mu(\bs)>0; \mathcal{B} \right)$ in $D1$ and $D5$   in both regions.  This plot   confirms the shift in location of the distribution  of the extent in the two regions.  Further, it shows  that the variability of the extent in $\mathcal{B}1$ is slightly lower than in $\mathcal{B}2$, in both decades.   Figure \ref{F6c} (left) shows the posterior density of     $\widetilde{Ext}\left(\bar Y_{D5}(\bs)-\bar Y_{D1}(\bs) > c;\mathcal{B} \right)$ for $c=0,1,2^\circ$C    in both regions. 
 
\bibliographystyle{copernicus}       
\bibliography{template.bib}   